\documentclass[11pt, a4paper,twoside,reqno]{amsart}

\makeatletter
\def\@settitle{\begin{center}%
		\baselineskip14\p@\relax
		\normalfont\LARGE\scshape\bfseries
		\@title
	\end{center}%
}

\def\@setauthors{%
  \begingroup
  \def\thanks{\protect\thanks@warning}%
  \trivlist
  \centering\small \@topsep30\p@\relax
  \advance\@topsep by -\baselineskip
  \item\relax
  \author@andify\authors
  \def\\{\protect\linebreak}%
  \authors%
  \ifx\@empty\contribs
  \else
    ,\penalty-3 \space \@setcontribs
    \@closetoccontribs
  \fi
  \endtrivlist
  \endgroup
}

\makeatother

\makeatletter

\def\subsection{\@startsection{subsection}{2}%
	\z@{.5\linespacing\@plus.7\linespacing}{.5\linespacing}%
	{\normalfont\large\bfseries}}

\def\subsubsection{\@startsection{subsubsection}{3}%
	\z@{.5\linespacing\@plus.7\linespacing}{.5\linespacing}%
	{\normalfont\itshape}}

\usepackage[usenames, dvipsnames]{xcolor}
\definecolor{darkblue}{rgb}{0.0, 0.0, 0.45}
\usepackage[hidelinks]{hyperref}
\hypersetup{colorlinks	= true,
raiselinks	= false,
linkcolor	= darkblue, 
citecolor	= Mahogany,
urlcolor	= ForestGreen,
pdfauthor	= {Mohammad Boveiri},
pdftitle	= {},
pdfkeywords	= {},
pdfsubject	= {},
plainpages	= false}

\allowdisplaybreaks
\date{\today}

\usepackage{amsthm}
\usepackage[english]{babel}
\usepackage[utf8]{inputenc} 
\usepackage[T1]{fontenc}    
\usepackage{hyperref}       
\usepackage{url}            
\usepackage{booktabs}       
\usepackage{amsfonts}       
\usepackage{nicefrac}       
\usepackage{microtype}      
\usepackage{appendix} 
\usepackage{lipsum}		
\usepackage{graphicx}
\usepackage{doi}
\usepackage{amsmath,amssymb,amsfonts}
\usepackage{comment}
\usepackage{subcaption}
\usepackage{biblatex}
\newcommand{\sign}{\mathrm{sign}}
\newcommand*{\argmin}{\operatornamewithlimits{argmin}\limits}
\newtheorem{theorem}{Theorem}
    \newtheorem{definition}{Definition}
    \newtheorem{proposition}[theorem]{Proposition}
    
    \newtheorem{lemma}[theorem]{Lemma}
    \newtheorem{remark}{Remark}
    \newtheorem{example}{Example}
    \newtheorem*{example*}{Example}

\title{Accelerating Adaptive Systems via \\ Normalized Parameter Estimation Laws}%
\author{Mohammad Boveiri$^{\text{a}}$, Peyman {Mohajerin Esfahani}$^{\text{a,b}}$, Mohammad Khosravi$^{\text{a}}$}
\thanks{The authors are with (a) Delft University of Technology, The Netherlands; and (b) University of Toronto, Canada. This work was partially supported by the ERC grant TRUST-949796.
}

\date{}

\begin{document}
\maketitle

\begin{abstract}
In this paper, we propose a new class of parameter estimation laws for adaptive systems, called \emph{normalized parameter estimation laws}. 
A key feature of these estimation laws is that they accelerate the convergence of the system state, $\mathit{x(t)}$, to the origin. 
We quantify this improvement by showing that our estimation laws guarantee finite integrability of the $\mathit{r}$-th root of the squared norm of the system state, i.e.,
\(
\mathit{\|x(t)\|}_2^{2/\mathit{r}} \in \mathcal{L}_1,
\)
where $\mathit{r} \geq 1$ is a pre-specified parameter that, for a broad class of systems, can be chosen arbitrarily large. 
In contrast, standard Lyapunov-based estimation laws only guarantee integrability of $\mathit{\|x(t)\|}_2^2$ (i.e., $\mathit{r} = 1$). 
We motivate our method by showing that, for large values of $r$, this guarantee serves as a sparsity-promoting mechanism in the time domain, meaning that it penalizes prolonged signal duration and slow decay, thereby promoting faster convergence of $\mathit{x(t)}$. 
The proposed estimation laws do not rely on time-varying or high adaptation gains and do not require persistent excitation. 
Moreover, they can be applied to systems with matched and unmatched uncertainties, regardless of their dynamic structure, as long as a control Lyapunov function (CLF) exists. 
Finally, they are compatible with any CLF-based certainty equivalence controllers.
We further develop higher-order extensions of our estimation laws by incorporating momentum into the estimation dynamics. 
We illustrate the performance improvements achieved with the proposed scheme through various numerical experiments.
\end{abstract}


\section{Introduction}
control is a powerful methodology that enables real-time learning, i.e., model parameter estimation, and performance improvement in dynamic systems while ensuring stability. Due to this key feature, it has been studied for decades \cite{ioannou1996robust,Krstic,slotine1991applied}, with applications across diverse fields, including robotics \cite{slotine1987adaptive,Spong,ortega1998passivity,spong.2022}, aerospace engineering \cite{flight.1,flight.2}, automated driving~\cite{taylor,driving}, and energy systems~\cite{Power.1,ortega1998passivity}.

The primary objective in adaptive control is the convergence of the system state to a desired value or trajectory, whereas the parameter convergence, while desirable, is of secondary importance.
Parameter convergence occurs only under a strong condition known as persistent excitation (PE) \cite {Krstic,slotine1991applied}, which typically arises from sufficiently rich reference inputs.

In the presence of PE, the system state and parameter estimate converge respectively to their desired and actual values with a fast rate, typically exponential \cite{ioannou1996robust,bastin2013line,SLOTINE.com,Mazenc}. Nonetheless,  in most control tasks, such as stabilization and set-point regulation, the PE property is generally absent. 
In standard Lyapunov-based adaptive control design, without the PE condition, the parameter estimate does not converge to the true value, and the convergence of the system state can be slow.
The primary reason for slow convergence in the absence of PE is that parameter adaptation decelerates as the system state approaches the origin. Typically, in this scenario, it can only be shown that the squared norm of the system state asymptotically converges to zero and belongs to 
the space of integrable functions. The reader is referred to Section \ref{sec:moti} for a detailed discussion of this topic.

\textbf{Related works.}
Research on accelerating the convergence of adaptive systems can be broadly divided into the following categories.

(i) \textbf{Convergence via altering the reference signal.
} A common approach to accelerate convergence is to enrich the reference signal, for instance by adding time-varying noise, thereby trying to enforce PE and accelerate the convergence of both the parameter estimate and the system state~\cite{boyd_sastry_1986,narendra1987persistent}. However, this comes at the cost of compromising the transient behavior of the system. Moreover, if the PE of the reference signal does not imply the PE of the regressor vector, this approach does not necessarily accelerate convergence~\cite{jenkins}.

(ii) \textbf{Parameter estimator design for a given controller.} This approach, which is conceptually closer to our work, focuses on parameter estimation algorithms without modifying the control law. 
In line with recent advances in optimization and machine learning \cite{bach2024learning,nesterov2004introductory,nemirovski2009robust}, this line of research incorporates ideas such as natural gradient and mirror descent-like algorithms \cite{Boffi.1,lee2018natural,wensing2018beyond}, as well as momentum-based methods \cite{Gaudio}, to exploit the geometry of the problem and improve convergence. However, to the best of our knowledge, the improvements are primarily empirical. More precisely, the theoretical guarantees remain similar to those of standard Lyapunov-based parameter estimation algorithms; namely, asymptotic convergence of the error to zero, with the squared error signal belonging to the space of
integrable functions.

(iii) \textbf{Joint controller and parameter estimator design.} This category includes works that design new adaptive controllers by modifying both the parameter estimation law and the control law, mainly for specific classes of nonlinear systems. For instance, the method in \cite{Exp, SONG1992271} applies a speed transformation to construct an accelerated (virtual) system, whose stabilization enables arbitrary adjustment of the convergence rate in the original system. This approach ensures exponential convergence without requiring persistent excitation. However, it relies on time-varying adaptation and control gains that grow exponentially over time, and is only applicable to certain system classes, such as those in strict-feedback form. Similarly, the works in \cite{Bech.1, Bech.2} propose adaptive controllers for feedback-linearizable and strict-feedback systems, guaranteeing convergence to a predefined residual set at a prescribed rate. However, this guarantee is non-asymptotic, i.e., the specified convergence rate does not apply when the error signal is close to zero, namely within the residual set. Adaptive fixed-time stabilization \cite{Meng-fix,WANG2020104704} is also proposed for a specific class of nonlinear systems, such as nonlinear systems with linear growth conditions.

\textbf{Contributions.} This study conceptually falls into category (ii), where we propose a new class of parameter estimation laws, called \emph{normalized parameter estimation laws}, designed to accelerate convergence in adaptive systems. The introduced estimation laws incorporate a form of normalization based on the Lyapunov function, enabling faster adaptation when the system state is close to the origin. We note that our parameter update laws do not rely on persistent excitation, time-varying adaptation gains, or side information. Additionally, they have the following distinct features:

(i) \textbf{Promoting sparsity in time-domain: 
Penalizing signal duration and slow decay.} 
We introduce the notion of the \emph{vanishing degree} $vd$ (Definition 2) for a given nonlinear system, an intrinsic property of a system that characterizes its behavior around the origin. 
We show that for any $r \leq vd$, our estimation laws ensure $r$th-root integrability of the squared norm of the system state, i.e., $\|x(t)\|_2^{2/r} \in \mathcal{L}_1$ (Theorem \ref{theorem_main}). 
We demonstrate that, for large values of $r$, this guarantee serves as a form of sparsity-promoting mechanism in the time domain, meaning it penalizes signal duration and slow decay of $x(t)$, thereby accelerating convergence of the system state. This effect becomes more pronounced as $r$ increases.
\par

Further, we show that for a considerably large class of nonlinear systems, namely systems that are exponentially stabilizable and have a regressor that is zero at the origin, the vanishing degree $vd$ is unbounded (Theorem \ref{theorem.vanish}), allowing the parameter $r$ to be chosen arbitrarily large.

(ii) \textbf{Compatibility with certainty-equivalence CLF-based controllers.} Our parameter estimation laws are applicable to systems with both matched and unmatched uncertainties. Moreover, they can be seamlessly integrated into any certainty-equivalence Lyapunov-based control scheme.\par
(iii) \textbf{Extension to higher-order estimation laws.} We further develop higher-order extensions of our estimation laws by incorporating momentum into the update dynamics (Theorem~\ref{thm:momentum}). We prove that these momentum-based algorithms are stable and globally convergent.

\textbf{Structure of the paper.} The rest of this paper is organized as follows. Section \ref{sec:moti} presents a motivating example of a scalar system, illustrating how our proposed parameter estimation law accelerates convergence. Section \ref{sec:main} generalizes this result to nonlinear systems with unmatched uncertainty. In Section \ref{sec:moment} we propose a higher-order extension of our estimation law incorporating momentum. Section \ref{sec:Proofs} provides the technical proofs of theorems and propositions presented in the paper. Simulation results in Section \ref{sec:sim} demonstrate the performance improvements achieved with the proposed method. Finally, in Section \ref{sec:dis}, we discuss some conclusions and future directions.\par

\textbf{Notation.} Throughout the paper, 
\(\mathbb{R}\) and \(\mathbb{R}_{\geq 0}\) denote the set of real numbers and non-negative real numbers, respectively. The symbol \( \mathcal{L}_p \) denotes the set of functions \( z : \mathbb{R}_{\geq0} \to \mathbb{R}^n\) such that $
 \int_0^\infty \|z(t)\|_2^p \,\mathrm{d}t < \infty.
$ 
Given a real-valued function \( V : \mathbb{R}^n \to \mathbb{R} \), we denote the vector of partial derivatives by $
\frac{\partial V(x)}{\partial x} = \left[\frac{\partial V}{\partial x_1}, \dots, \frac{\partial V}{\partial x_s}\right]$. A function $V$ is said to be of class $C^r$ if all its partial derivatives up to order $r$ exist and are continuous. Furthermore, a function is called smooth, or of class $C^\infty$, if its partial derivatives of all orders exist and are continuous.
A function \( V(x) \) is positive definite if \( V(0) = 0 \) and \( V(x) > 0 \) for all \( x \neq \mathbf{0} \). Moreover, \( V(x) \) is radially unbounded if \( V(x) \to \infty \) as \( \|x\| \to \infty \). The span of a set of vectors $\{v_1, \dots, v_m\}$ is given by
\begin{align*}
\operatorname{span}\{v_1, \dots, v_m\}
    = \{\, v \mid \exists \lambda_1, \dots, \lambda_m \in \mathbb{R}, \quad \text{s.t. } v = \lambda_1 v_1 + \cdots + \lambda_m v_m \,\}.
\end{align*}
For symmetric matrices \( A \) and \( B \), the notation \( A < B \) (resp. \( A \leq B \)) means that \( A - B \) is negative definite (resp. negative semi-definite).  
Furthermore, $I_{n}$ denotes the identity matrix of size $n\in\mathbb{N}$.

\section{ A Motivating Example and Preliminary Result}\label{sec:moti}

In this section, a motivating example is provided to briefly demonstrate the core idea of this article, namely, how suitable Lyapunov-based normalization of the parameter estimation law accelerates the convergence of the system state to the origin. To this end, we consider the scalar system described by the differential equation
\begin{equation}\label{eq:ex.dynamic.main}
\dot{x} = \theta |x| + u,
\end{equation}
where $x$ is the state, $u$ is the input, and $\theta$ is an unknown parameter. The goal is to design an adaptive controller to regulate the state to the origin. 
To better understand our proposed method and enable direct comparison, we begin by reviewing the standard direct adaptive control design procedure based on Lyapunov theory and elaborate on how it forms the foundation of our approach.

\vspace{-0.4cm}
\subsection{Standard Lyapunov-based estimation law.} Let $\hat{\theta}(t) \in \mathbb{R}$ denote an estimate of the unknown parameter. Consider the following augmented Lyapunov function
$$
Q(x, \tilde{\theta}) := V(x) + \frac{1}{2\gamma} \tilde{\theta}^2,
$$
where $\gamma> 0$, $\tilde{\theta} = \hat{\theta}(t) - \theta$ represents the parameter estimation error, and $V(x) := \frac{1}{2} x^2$ is a CLF for system \eqref{eq:ex.dynamic.main}, as detailed in Definition~\ref{def:eq.CCLF}. Taking the time derivative of $Q(x,\hat{\theta})$ along \eqref{eq:ex.dynamic.main} yields
\begin{equation*}
\dot{Q}(x, \tilde{\theta}) = \frac{\partial V(x)}{\partial x}\Big(\hat{\theta}|x| + u -\tilde{\theta} |x| \Big) +  \frac{1}{\gamma}\tilde{\theta} \dot{\hat{\theta}}.
\end{equation*}
By selecting the control law 
\begin{equation} \label{Eq:ex.control}
    u=-\hat{\theta}|x| - \frac{1}{2}x,
\end{equation}
and estimation law
\begin{equation}\label{EQ:ex.update.a}
    \dot{\hat{\theta}}=\gamma \frac{\partial V(x)}{\partial x} |x|=\gamma\, 
    \sign(x)\, x^2,
\end{equation}
we obtain $\dot{Q} 
   \le - V(x)  \leq 0$. Thus, $Q(x,\hat{\theta})$ is non-increasing, which implies that $x(t)$ and $\hat{\theta}(t)$ are bounded. Moreover, by integrating both sides with respect to time, we obtain
$
    \int_{0}^\infty V(x) \mathrm{d}t \leq Q\big(x(0), \hat{\theta}(0)\big) - Q\big(x(\infty), \hat{\theta}(\infty)\big) \leq Q\big(x(0), \hat{\theta}(0)\big) < \infty,
$
implying that \(V(x)\) is integrable, i.e., \(V(x) \in \mathcal{L}_1\). Thus, $x(t)\in \mathcal{L}_2$. Furthermore, we have 
$$
 \dot{V}(x) = -V(x)- \tilde{\theta}\,\sign(x)\, x^2,
$$
which shows that \(\dot{V}(x)\) is also bounded. As a result, \(V(x)\) is uniformly continuous. 
Barbalat’s Lemma, presented below, then guarantees that \(V(x)\) convergence to zero as 
$t$ goes to infinity, given its uniform continuity and integrability.

\begin{lemma}[Barbalat's Lemma \cite{Krstic,slotine1991applied}]
    Let \(v(t)\in\mathcal{L}_1\) be a uniformly continuous function defined on $[0,\infty)$.
    Then, $\displaystyle\lim_{t\to\infty}v(t)=0$.
\end{lemma}
Note that if the system parameter $\theta$ is exactly known and we set $\hat{\theta} = \theta$, then the controller~\eqref{Eq:ex.control} ensures exponential convergence of the system state at a fixed rate, since $\tilde{\theta} = 0$. However, when $\theta$ is unknown, we can generally only establish the asymptotic convergence of $x(t)$ to zero and its square integrability, i.e., $x(t) \in \mathcal{L}_2$. The primary reason for the slower convergence in the adaptive case is that, as $x(t)$ approaches zero, the estimation law slows down significantly, leading to slower parameter adaptation and, consequently, slower convergence of the system state.
\begin{figure}[t]
    \centering
    \begin{subfigure}{0.32\columnwidth}
        \centering
        \includegraphics[width=\textwidth]{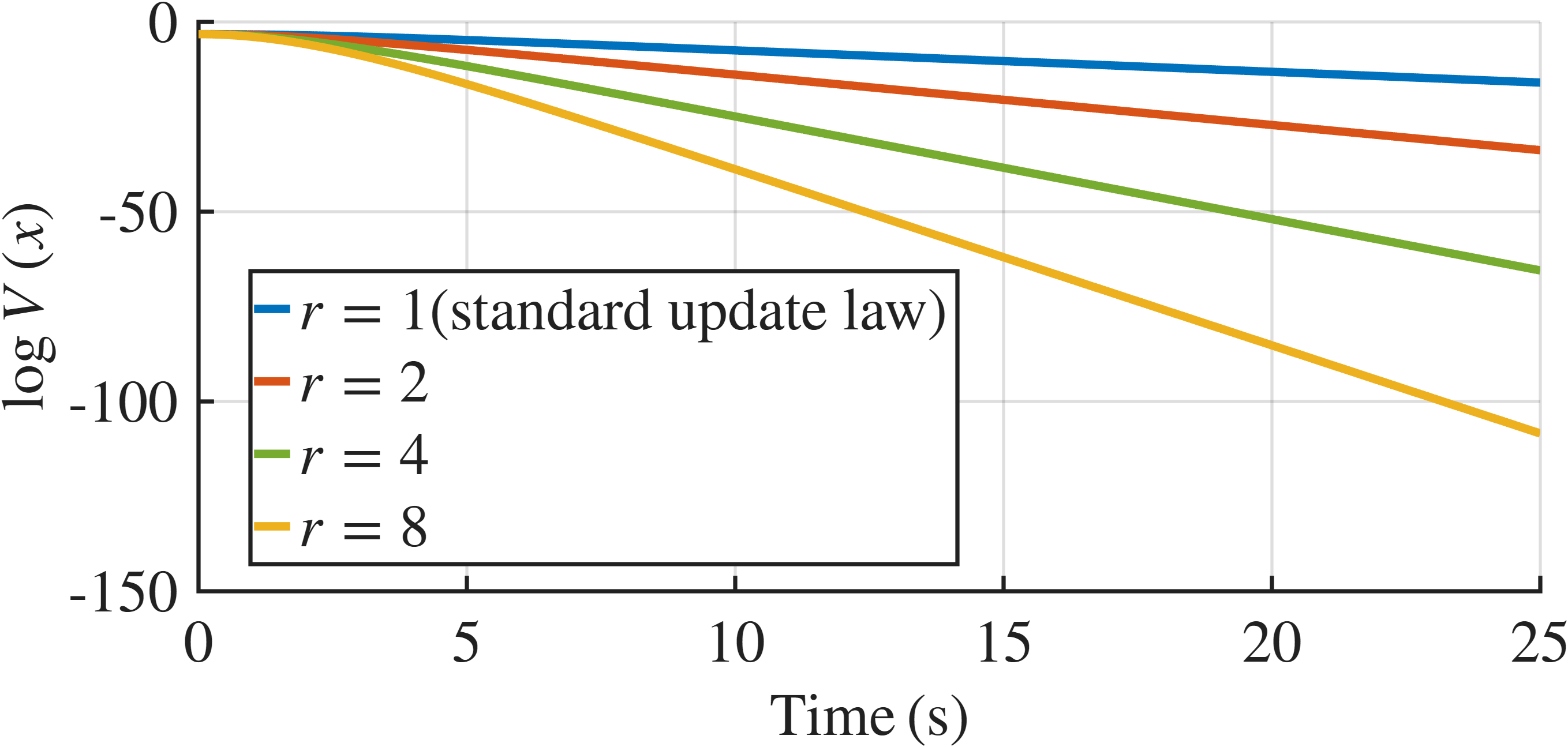}
        \caption{Evolution of $\log V(x)$}
           \label{fig:ex.1.logv}
    \end{subfigure}
    \hfill
    \begin{subfigure}{0.33\columnwidth}
        \centering
        \includegraphics[width=\textwidth]{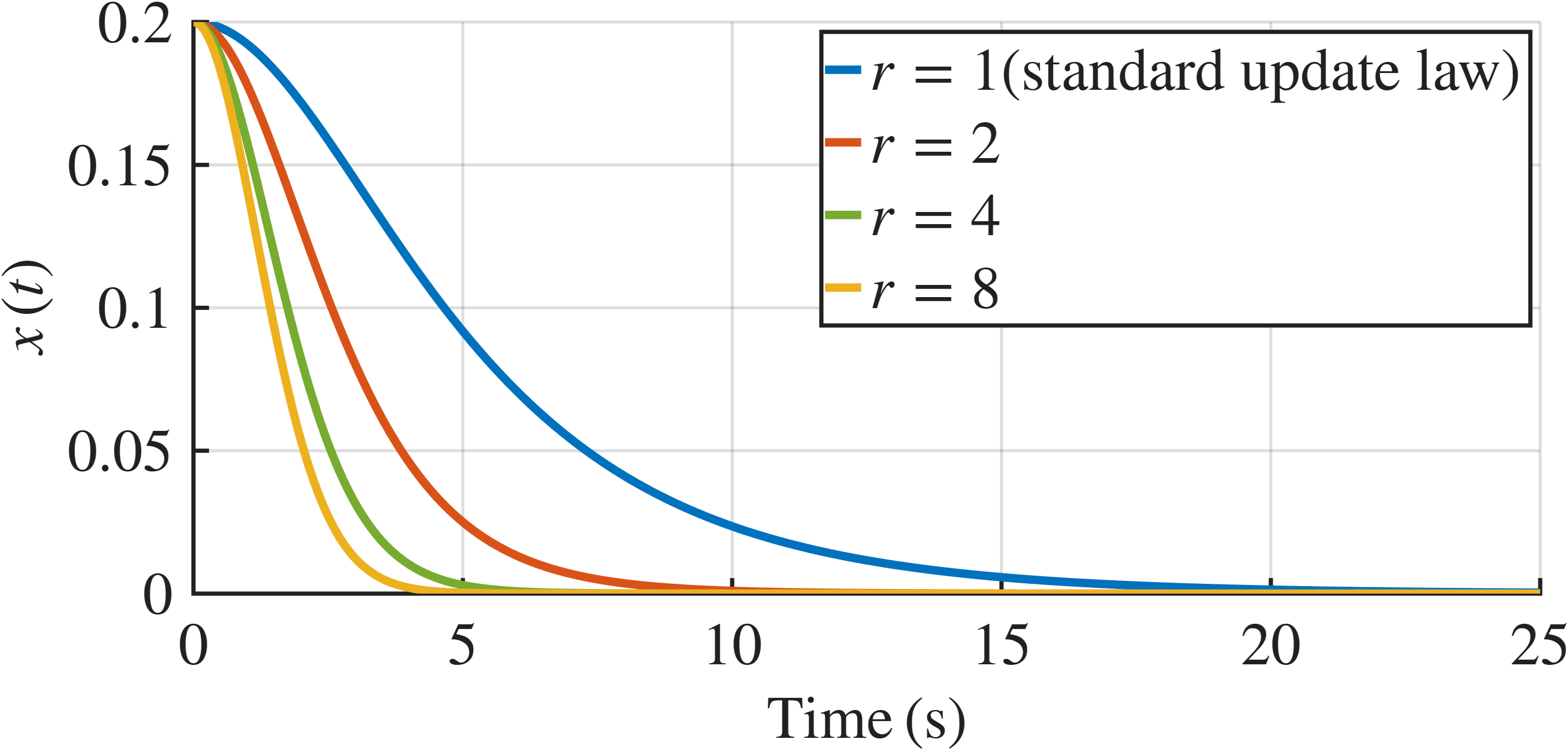}
        \caption{\( x(t) \) for different values of \( r \).}
             \label{fig:ex.1}
    \end{subfigure}
    \hfill
    \begin{subfigure}{0.33\columnwidth}
        \centering
        \includegraphics[width=\textwidth]{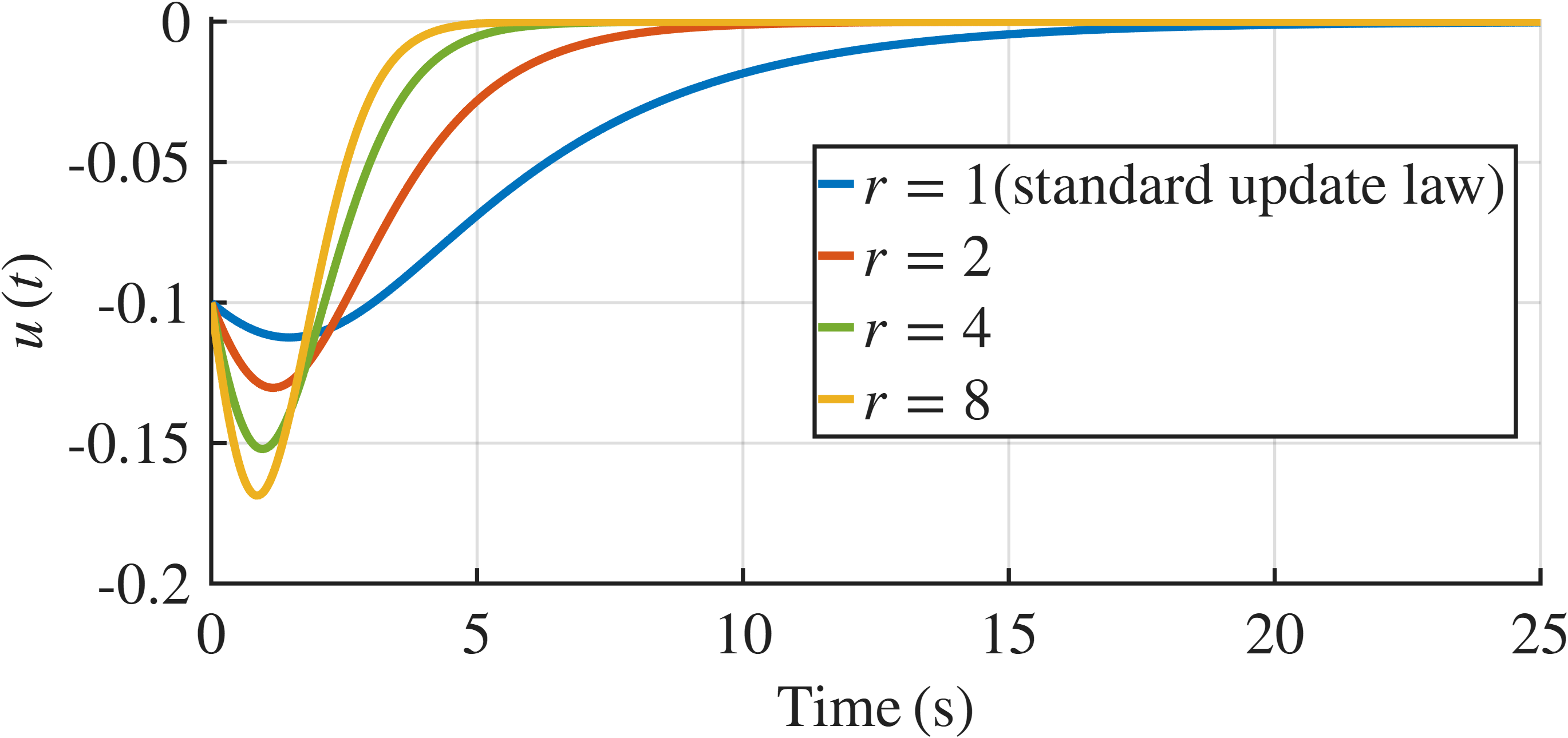}
        \caption{$u(t)$ for different values of  \( r \).}
              \label{fig:ex.1.ut}
    \end{subfigure}

       \caption{(a) Plots of the logarithm of the Lyapunov function 
\( V(x) = \tfrac{1}{2} x^2 \) versus time for 
system~\eqref{eq:ex.dynamic.main} under control law~\eqref{Eq:ex.control} 
and adaptive law~\eqref{eq.update.pro} with \( \gamma = 1 \) and different 
values of \( r \). 
(b) Closed-loop system state variable. 
(c) Control signal \( u(t) \) for different values of \( r \).}
\end{figure}

\subsection{ Normalized parameter estimation law} 
In this subsection, we provide an alternative parameter estimation law for system \eqref{eq:ex.dynamic.main} that accelerates the convergence of the system state to zero. 

\begin{proposition}[Normalized estimation law for \eqref{eq:ex.dynamic.main}]\label{proposition.1}
Consider the nonlinear system~\eqref{eq:ex.dynamic.main} with the control law~\eqref{Eq:ex.control} and the estimation law
\begin{equation}\label{eq.update.pro}
   \dot{\hat{\theta}} = \gamma\,\mathrm{sign}(x)\big( x^2 + |x|^{2/r} \big), 
\end{equation}
where $r \geq 1$.  
Then  $x(t)$ asymptotically converges to zero. Moreover, we have $V(x)^{1/r} \in \mathcal{L}_1$.
\end{proposition}
The proof of Proposition~\ref{proposition.1} is provided in Section~\ref{sec:appen:prop.1}.

\begin{remark}[Standard vs. normalized estimation law]\normalfont
A direct comparison between the standard estimation law~\eqref{EQ:ex.update.a} and the estimation law~\eqref{eq.update.pro} reveals the presence of an additional term of the form \( |x|^{2/r} \) in~\eqref{eq.update.pro}. This term prevents the adaptation from slowing down too quickly as the system state converges to zero, resulting in faster parameter adjustment and improved convergence of the system state. This effect becomes more pronounced as the parameter \(r\) increases. We illustrate this behavior in Figure~2, which corresponds to simulations where the control law~\eqref{Eq:ex.control} and estimation  law~\eqref{eq.update.pro} are applied to system~\eqref{eq:ex.dynamic.main}. Figure~\ref{fig:ex.1.logv} shows the logarithm of \( V(x) = \tfrac{1}{2}x^2 \) as a function of time, Figure~\ref{fig:ex.1} shows the evolution of the system state, and Figure~\ref{fig:ex.1.ut} shows the control signal \( u(t) \) applied to the system. We observe that increasing \( r \) accelerates the convergence of the system state to the origin.
\end{remark}

A key feature of the update law \eqref{eq.update.pro} is that it ensures $r$-th root integrability of $V(x)$, that is $V(x)^{1/r} \in \mathcal{L}_1$, rather than absolute integrability, i.e.,  $V(x) \in \mathcal{L}_1$, established in the previous section. We next outline the implications and properties of this guarantee.

\textbf{Monotonicity.} Let $\nu(t)$ be a nonnegative function defined on $[0,\infty)$. For any $r_1 \ge r_2 \ge 1$, if $\nu(t)^{1/r_1} \in \mathcal{L}_1$, then $\nu(t)^{1/r_2} \in \mathcal{L}_1$. In other words, the condition $\nu(t)^{1/r} \in \mathcal{L}_1$ becomes stronger as $r$ increases. To build intuition, consider $\nu(t) = \frac{1}{(t+1)^{q}}$ with $q \ge 1$. We have $\nu(t)^{1/r} \in \mathcal{L}_1$ if and only if $q > r$. When $q \leq r$, the decay of $\nu(t)$ is too slow and the integral diverges.

\textbf{Promoting sparsity: Penalizing signal duration for large $r$.} 
As $r \to \infty$, we have $\nu(t)^{1/r} \to \mathbf{1}_{\{\nu(t)\neq 0\}}$ where $\mathbf{1}_{\{\nu(t)\neq 0\}}$ is the indicator function defined as
\begin{equation*}
    \mathbf{1}_{\{\nu(t)\neq 0\}} \;=\;
    \begin{cases}
        1, & \nu(t)\neq 0, \\[6pt]
        0, & \nu(t)=0.
    \end{cases}
\end{equation*}
Hence, the integral
$$\int_0^\infty \mathbf{1}_{\{\nu(t)\neq 0\}} \mathrm{d}t=\int_0^\infty \lim_{r \to \infty } \nu(t)^{1/r} \mathrm{d}t $$
measures the length of the time interval where $\nu(t)$ is nonzero. Therefore, if this integral is bounded, it means that $\nu(t)$ is nonzero only for a finite amount of time.

Now consider the case where $\nu(t) = V(x(t)) = \frac{1}{2}x(t)^2$. While Proposition~\ref{proposition.1} does not guarantee the integrability of $V(x)^{1/r}$ as $r \to \infty$, it does guarantee the integrability of $V(x)^{1/r}$ for any finite $r$. Hence, increasing $r$ can be viewed as a sparsity-promoting mechanism as it penalizes signals with long duration and slow decay.
It is interesting to note that, as $r$ increases, the update law \eqref{eq.update.pro} approaches the limiting form
\begin{equation*}
     \lim_{r \to \infty} \dot{\hat{\theta}} \;=\; \gamma\,\mathrm{sign}(x)\Big( x^{2} + \mathbf{1}_{\{x(t)\neq 0\}} \Big). 
\end{equation*}
This estimation law maintains a constant adaptation rate even when the system state is close to zero, and switches off adaptation only when the state is exactly zero. 

\begin{example}[Estimation laws with smooth limits]\normalfont
The limiting forms of the normalized estimation laws (i.e., as $r \to \infty$) depend on the system dynamics and can be smooth. For instance, consider the system
\(
\dot{x} = \theta x^2 + u,
\)
with the control and estimation laws given by  
\[
u = -\hat{\theta}x^2 - \tfrac{1}{2}x, \quad 
\dot{\hat{\theta}} = \gamma \big(x^3 + x\,|x|^{2/r}\big),
\]
respectively.  
Following a similar line of reasoning as in the proof of Proposition~\ref{proposition.1}, one can show that $x(t) \to 0$ as $t \to \infty$, and $\|x(t)\|_2^{2/r} \in \mathcal{L}_1$.  
Moreover, the limiting form of the estimation law is  
\[
\lim_{r \to \infty} \dot{\hat{\theta}} = \gamma(x^3 + x),
\]
which is smooth.
\end{example}

\section{Accelerating Adaptation via Normalized Parameter Estimation Laws}
\label{sec:main}
In this section, we generalize the results and ideas from the previous section to the case of multidimensional nonlinear dynamical systems. 
In our analysis, we consider both matched and unmatched uncertainties. 

Consider an uncertain dynamical system of the form  
\begin{equation} \label{Eq.dynamic}  
    \dot{x} = f(x) + \Delta(x)^\top \theta + B(x) u,  
\end{equation}  
where $x \in \mathbb{R}^n$ is the system state, $u \in \mathbb{R}^m$ is the control input, $\theta \in \Theta \subseteq \mathbb{R}^p$ is an unknown vector of fixed parameters characterizing the uncertainty in the dynamics of the system, $\Theta$ is a convex set, and  
\( f : \mathbb{R}^n \rightarrow \mathbb{R}^n \), 
\( B : \mathbb{R}^n \rightarrow \mathbb{R}^{n \times m} \), and \( \Delta : \mathbb{R}^n \rightarrow \mathbb{R}^{p \times n} \) are known functions. For ease of discussion, we focus here on the stabilization problem of the above system, i.e., driving the system state to the origin, $x=0$. In such problems, persistent excitation is typically absent due to the lack of richness of the input signal.

We begin by recalling the definition of control Lyapunov functions for the introduced system (see, e.g., \cite{khalil2002nonlinear, Krstic})
\begin{definition}[Control Lyapunov function]\label{def:eq.CCLF}
A smooth and positive definite function $V(x,\theta)$ is an exponential control Lyapunov function (CLF) for \eqref{Eq.dynamic} if there exist positive constants $c_1$, $c_2$, and $\lambda$ such that
\begin{align}
&\quad\ \, 
c_1 \|x\|_2^2\leq V(x,\theta) \leq c_2 \|x\|_2^2, \label{eq.clf.a}\\
&
       \inf_{u \in \mathbb{R}^m} \frac{\partial V(x,\theta) }{\partial x}\left(f(x)+\Delta(x)^\top\theta+
       B(x)u\right)\leq -\lambda V(x,\theta),
       \label{eq.clf}
\end{align}
for all $x\in\mathbb{R}^n$ and $\theta\in\Theta$.
\end{definition}
We note that the existence of a CLF is a necessary and sufficient condition for the exponential stabilizability of a dynamical system \cite[Theorem 4.14]{khalil2002nonlinear}. Depending on the structure of the system,  CLFs can be constructed using various methods, including physics-based approaches \cite{slotine1987adaptive, Spong, ortega1998passivity}, numerical techniques, such as semidefinite and sum-of-squares programming \cite{boyd1994linear, Sos}, and control design tools, like back-stepping and feedback linearization \cite{Krstic, freeman2008robust}.
 
\textbf{CLF-based Controllers.} 
Let \( V(x,\theta) \) be a CLF for the nonlinear system described by \eqref{Eq.dynamic}, for example, obtained through the aforementioned methodologies.
Let $\hat{\theta}(t)$ be an estimate of the unknown parameter $\theta$ at time $t$, and assume that a controller $u(t)$ is in place such that the inequality 
\begin{equation}\label{eq.C}  
  \frac{\partial V(x,\hat{\theta})}{\partial x} \Big(f(x) + \Delta(x)^\top \hat{\theta} + B(x) u\Big) \leq -\lambda V(x,\hat{\theta})
\end{equation}  
holds.
Note that if $\hat{\theta}(t) = \theta$, i.e., when the estimated parameter vector exactly matches the true system parameter vector, then the above controller ensures that $V(x, \hat{\theta})$ converges to zero exponentially. Such controllers are referred to as \emph{certainty-equivalence controllers} \cite{Krstic}, as they are derived by simply replacing $\theta$ with $\hat{\theta}$.
%
There are several well-known controllers that satisfy the inequality \eqref{eq.C}. One example is the point-wise minimum norm controller, defined by
\begin{equation}\label{eq.theore.C}
u(x,\hat{\theta}) = \displaystyle\argmin_{\bar{u} \in \mathbb{R}^m} \left\{ \bar{u}^\top \bar{u} \;\middle|\; \text{subject to } \eqref{eq.C} \right\}.
\end{equation}

\textbf{Parameter estimation law.} Similar to Section~\ref{sec:moti}, our main objective is to develop an update law for the parameter estimate $\hat{\theta}$ in \eqref{eq.C} that guarantees finite integrability of $\|x\|_2^{2/r}$ for a prespecified constant $r\in \mathbb{R}_{>0}$. Achieving this requires characterizing how the regressor $\Delta(x)$ converges to zero with respect to a given CLF. To this end, we introduce the notion of the \emph{vanishing degree} of the regressor $\Delta(x)$, which is an intrinsic attribute of the system.

\begin{definition}[Vanishing degree]
Given the CLF \(V(x,\theta)\), we define \emph{vanishing degree}, denoted by $vd$, of the regressor $\Delta(x)$ with respect to $V$ as
\begin{equation*}\!\!\!\!
vd(\Delta,V) \,:=\,
\sup\Bigl\{r \in \mathbb{R}_{>0} \;\Big|\;\
    \lim_{x \to 0}
      \big\|
        V(x,\theta)^{1/r - 1}\,\frac{ \partial V(x,\theta)}{\partial x}\,\Delta(x)^\top\, 
      \big\|_{\infty}
    < \infty, \; \forall\theta \in \Theta \Big\}.
\end{equation*}
\end{definition}

The following example further illustrates the notion of vanishing degree discussed above.
\begin{example}[Vanishing degree] \normalfont
Consider the scalar system
$
\dot{x} = \Delta(x)^\top \theta + u,    
$
with
$
\Delta(x) = [\sin(x),\; x,\; x^2]^\top,
$
and the CLF $V(x) = x^2$. For this system,  we have

\begin{equation*}
    \lim_{x \to 0}
        V(x)^{ 1/r - 1}\,\frac{ \partial V(x)}{\partial x}\,\Delta(x)^\top\, 
      = \lim_{x \to 0}
    \,|x|^{2/r}
   \,\,\, [\frac{\sin(x)}{x},\; 1,\; x]=0,
\end{equation*}

for all $r\in \mathbb{R}_{>0}$. Therefore, the vanishing degree of this system is unbounded, i.e., $vd(\Delta,V) = \infty$. 
We have a similar property for higher-dimensional systems, provided that  \(\Delta(x) \) approaches zero at least linearly as \(x\) goes to zero (see Theorem \ref{theorem.vanish} for more details). 
On the other hand, if \(\Delta(x)\) includes any component that is nonzero at the origin, such as \(\cos(x)\) or a nonzero constant, then  \(vd(\Delta,V) = 2\).
\end{example}

The following theorem establishes an update law for $\hat{\theta}(t)$ that ensures $\|x(t)\|^{2/r} \in \mathcal{L}_1$ for $r \leq vd(\Delta, V)$.

\begin{theorem}[Normalized parameter estimation law]\label{theorem_main} 
Let $\varepsilon > 0$ be a scalar, $\Gamma \in \mathbb{R}^{p \times p}$ a positive definite matrix, and $\omega \in C^1$ a continuously differentiable function satisfying $\inf_{\rho\in\mathbb{R}}\omega(\rho)>0$, and $\omega_\rho(\rho):= \tfrac{\partial \omega(\rho)}{\partial \rho} > 0$. 
Let $u(t)$ be any control signal satisfying \eqref{eq.C}, with the estimation law defined as
\begin{subequations}  
\begin{align}
\label{eq:update.theta}  
\dot{\hat{\theta}} &= \Gamma \, \omega(\rho)\, V(x,\hat{\theta})^{1/r - 1}\, \Delta(x)\,
\frac{\partial V(x,\hat{\theta})}{\partial x}^\top, 
\\[6pt]  
\label{eq:update.rho}  
\dot{\rho} &= - \frac{\omega(\rho)\,V(x,\hat{\theta})^{1/r - 1}}
{\omega_\rho(\rho)\,\big(r V(x,\hat{\theta})^{1/r} + \varepsilon \big)}\,
\frac{\partial V(x,\hat{\theta})}{\partial \hat{\theta}}\, \dot{\hat{\theta}},
\end{align}  
\end{subequations}
where $r$ is a positive real scalar such that $1 \leq r \leq vd(\Delta,V)$. 
Then the following properties hold:  
\begin{enumerate}
    \item All closed-loop signals remain bounded, and \(
\displaystyle \lim_{t \to \infty} x(t) = 0
\).  
    \item  The state convergence is accelerated in the sense that $\|x(t)\|_2^{2/r} \in \mathcal{L}_1$.
\end{enumerate}  

\end{theorem}
The proof of Theorem \ref{theorem_main} is provided in Section~\ref{apn:proof_theorem_main}.
\begin{remark}[Simpler update law for matched uncertainty]\label{remark.match}\normalfont
When the uncertainty satisfies the \emph{matching condition}
$
\Delta(x)^\top\theta \in \text{span}\{b_1(x), \ldots,b_m(x)\},
$
where $b_1(x), \ldots,b_m(x)$ are the columns of input matrix $B(x)$,
parameter-independent CLFs can be constructed for \eqref{Eq.dynamic}. 
Specifically, any CLF \(V(x)\) for the nominal system  
$
\dot{x} = f(x) + B(x)u,
$  
remains valid for the uncertain system \eqref{Eq.dynamic}, since the term \(\Delta(x)^\top \theta\) can be canceled by the control input. As a result, in the matched uncertainty case, the update law \eqref{eq:update.rho} becomes zero, i.e., the update law \eqref{eq:update.rho} is only active when the CLF depends on the system parameters. The main idea behind the update law \eqref{eq:update.rho}, inspired by \cite{Lopez.universal}, is to modulate the adaptation rate so that transients in parameter estimation do not compromise stability.
\end{remark}
\vspace{-0.5 cm}

\begin{remark}[Robust modifications]\normalfont
Since our framework is Lyapunov-based, standard robustness modifications from the literature can be seamlessly incorporated into both the update law and the control law to improve performance in the presence of disturbances and sensor noise. For the parameter update law, projection operators \cite[p.~511]{Krstic} can be used to incorporate prior bounds on the parameters, ensuring that the estimates remain within a predefined set. Alternatively, dead-zone techniques (see, e.g., \cite[Propositions 1 and 2]{lopez2019contraction}) can be employed to guarantee boundedness of the parameter estimates. For the control law, robustness to matched disturbances can be improved by incorporating nonlinear damping \cite[Lemma~2.26]{Krstic}.
\end{remark}

Note that the parameter $r$ in Theorem \ref{theorem_main} is bounded above by the vanishing degree. 
In the next theorem, we show that for a broad class of dynamical systems, the vanishing degree is infinite, i.e., $vd(\Delta, V) = \infty$. 
Consequently, for these systems, the parameter $r$ can be chosen arbitrarily large.

\begin{theorem} [Computing vanishing degree]\label{theorem.vanish} Let $V(x,\theta)$ be a CLF for system \eqref{Eq.dynamic}, and the regressor $\Delta(x)$ be twice continuously differentiable, i.e., $\Delta(x)\in C^2$, with $\Delta(0) = 0$. Then, the corresponding vanishing degree is infinite, i.e., $vd(\Delta,V) = \infty$.
\end{theorem}

The proof of Theorem~\ref{theorem.vanish} is provided in Section~\ref{sec:appen:thm.vanish}.  We note that the condition $\Delta(0) = 0$ mentioned in Theorem \ref{theorem.vanish} is common in adaptive stabilization of nonlinear systems, e.g., see \cite{Krstic,chen2021adaptive}.

\vspace{-0.25cm}
\section{Normalized estimation laws with momentum}\label{sec:moment}
In this section, we propose a new parameter estimation law for nonlinear systems with unmatched uncertainty, incorporating the normalization technique from the previous section along with a momentum term. The developed estimation law introduces second-order dynamics for the parameter estimate $\hat{\theta}(t)$, similar to the momentum-based algorithms commonly used in the optimization literature \cite{nesterov1983method,boyd2004convex}. We prove general conditions under which these momentum algorithms provide guarantees of the form $\| x(t)\|^{2/r} \in \mathcal{L}_1$ for any prespecified $r \geq 1$. While the results in this section can be stated more generally, for ease of discussion, we focus on the case where a quadratic CLF can be found for system \eqref{Eq.dynamic}. This assumption is closely related to the concept of quadratic stabilizability discussed in \cite{boyd1994linear,pavlov2006uniform}. The following theorem introduces a higher-order update law for $\hat{\theta}(t)$, which ensures $\|x(t)\|^{2/r} \in \mathcal{L}_1$ for any $r \geq 1$.

\begin{theorem}[Normalized estimation law with momentum] \label{thm:momentum} Suppose that
there exists a quadratic CLF of the form $V(x,\theta) = \frac{1}{2}x^\top P(\theta) x$ for system \eqref{Eq.dynamic}, and the regressor \( \Delta(x) \) satisfies the conditions stated in Theorem \ref{theorem.vanish}.
Also, let $\varepsilon$, $\Gamma$, and $\omega$ be specified as in Theorem~\ref{theorem_main}.
Then, for any control signal $u(t)$ satisfying \eqref{eq.C} and the estimation laws defined as
\begin{subequations}
\begin{align} \label{eq:update.a.col}
    \dot{\hat{a}} &= \Gamma\,  \omega(\rho)  V(x,\hat{\theta})^{1/r - 1}  \Delta(x)  \,
    \frac{\partial V(x,\hat{\theta})}{\partial x}^\top, \\  \label{eq:update.theta.col}  
    \dot{\hat{\theta}}&=\frac{4}{\lambda}\, \Gamma\,\omega(\rho)\,V(x,\hat{\theta})^{1/r - 1}\Delta(x)P(\hat{\theta})\Delta(x)^\top(\hat{a}-\hat{\theta}),\\
    \label{eq:update.rho.col}  
    \dot{\rho}  &= - \frac{\omega(\rho)V(x,\hat{\theta})^{1/r - 1}}{\omega_\rho(\rho)\Big(r V(x,\hat{\theta})^{1/r} + \varepsilon \Big)}\,\frac{\partial V(x,\hat{\theta})}{\partial \hat{\theta}} \dot{\hat{\theta}},  
\end{align}  
\end{subequations}
the resulting closed-loop signals remain bounded, and the system state $x(t)$ converges to zero as $t \to \infty$. Moreover, the state convergence is accelerated in the sense that~$\|x\|_2^{2/r} \in \mathcal{L}_1$.
\end{theorem}

The proof of Theorem~\ref{thm:momentum} is provided in Section~\ref{sec:appen:thm.momentum}. Similar to the update laws presented in Theorem~\ref{theorem_main}, when the uncertainty is matched, the update law~\eqref{eq:update.rho.col} becomes inactive, resulting in simpler update dynamics.
\begin{remark}[Unmatched uncertainty and fast convergence]
\normalfont
Compared to existing higher-order parameter estimation laws \cite{Boffi.1, Gaudio}, the estimation law proposed in Theorem~\ref{thm:momentum} offers two key advantages. First, it can handle unmatched uncertainties. Second, it guarantees that $\|x(t)\|_2^{2/r} \in \mathcal{L}_1$ for any $r \in \mathbb{R}_{>0}$, whereas the methods in \cite{Boffi.1, Gaudio} achieve only $\|x(t)\|_2^2 \in \mathcal{L}_1$ (i.e., the case $r = 1$).
\end{remark}

\section{Technical Proofs}\label{sec:Proofs}
Here, we provide the proofs of the theorems and propositions presented in the preceding sections.
\subsection{Proof of Proposition~\ref{proposition.1}}\label{sec:appen:prop.1}
Consider the following nonnegative function
$Q(x,\hat{\theta}):=x^2+r|x|^{2/r}+\frac{1}{\gamma}\tilde{\theta}^2.$
By taking the time derivative of $Q(x,\hat{\theta})$ along the dynamics \eqref{eq:ex.dynamic.main}, we find that
\begin{align*}
    \dot{Q}\,=\,2(x+\sign(x)|x|^{2/r-1})(\hat{\theta}|x|+u-\tilde{\theta}|x|) +\frac{2}{\gamma}\tilde{\theta}\dot{\hat{\theta}}
\end{align*}
Substituting the control law \eqref{Eq:ex.control}  and update law  \eqref{eq.update.pro} into the above inequality gives
$\dot{Q}= -x^2-|x|^{2/r}\leq 0$,  which means that $x(t)$ and $\hat{\theta}(t)$ are bounded. By integrating both sides of the above inequality with respect to time, we find that
$\int_0^\infty \! x^2(t)+ |x(t)|^{2/r}\,\mathrm{d}t 
    \leq 
    Q\big(x(0),\hat{\theta(0)\big)} <\infty$,
meaning that $|x(t)|^{2/r}$ is integrable, i.e., $|x(t)|^{2/r} \in \mathcal{L}_1$. Moreover, we have
\begin{align}
    \frac{\mathrm{d}}{\mathrm{d}t} r |x|^{2/r} 
    &\, 
    =  
    2 \sign(x) |x|^{2/r-1} (\hat{\theta} |x|+u)-2\tilde{\theta} \sign(x)|x|^{2/r} 
   \nonumber \\
    &\,
    =
    -|x|^{2/r}-2\tilde{\theta}\sign(x)|x|^{2/r}.\label{eqn:proof_prop1_4}
\end{align}
Since \( x(t) \) and \( \hat{\theta}(t) \) are bounded, the right-hand side of \eqref{eqn:proof_prop1_4} is  bounded. As a result, \( |x(t)|^{2/r} \) is uniformly continuous. Hence, by Barbalat's lemma, $x(t)$ converges to zero as $t \to \infty$.
\qed

\subsection{Proof of Theorem~\ref{theorem_main}}
\label{apn:proof_theorem_main} 
Let the function $Q:\mathbb{R}^n\times\mathbb{R}^p\times \mathbb{R}\to\mathbb{R}$ be defined as 
\begin{equation*}    
Q
\big(x,\hat{\theta},\rho\big)
=
\omega(\rho) \big( r {V(x,\hat{\theta})^{1/r}} +\varepsilon \big)+\frac{1}{2}\tilde{\theta}^\top\Gamma^{-1}\tilde{\theta},
\end{equation*}
Taking the time derivative of $Q$ along the dynamics~\eqref{Eq.dynamic} gives
\begin{align}
\dot{Q}=\,
&\omega(\rho)V(x,\hat{\theta})^{1/r-1} \frac{\partial V(x,\hat{\theta})}{\partial x} \Big(f(x) +\Delta(x)^\top \theta +B(x) u \Big) \nonumber+\omega(\rho) V(x,\hat{\theta})^{1/r-1} \frac{\partial V(x,\hat{\theta})}{\partial \theta}\dot{\hat{\theta}}\nonumber\\
&+ \frac{\partial \omega(\rho)}{\partial \rho} \dot{\rho}\big( r V(x,\hat{\theta})^{1/r} +\varepsilon \big)+ \tilde{\theta}^\top\Gamma^{-1}\dot{\hat{\theta}}.\label{eqn:proof_main_thm_Qdot_2nd_eq}
\end{align}
By substituting $u$ in right-hand sight of \eqref{eqn:proof_main_thm_Qdot_2nd_eq}  with a control law satisfying \eqref{eq.C} and
 rearranging the terms, it follows that
\begin{align}
\dot{Q}\,\leq\ 
& -\omega(\rho)\lambda V(x,\hat{\theta})^{1/r}
-\omega(\rho)V(x,\hat{\theta})^{1/r-1} \frac{\partial V(x,\hat{\theta})}{\partial x} \nonumber\Delta(x)^\top\tilde{\theta} 
\nonumber+\omega(\rho) V(x,\hat{\theta})^{1/r-1} \frac{\partial V(x,\hat{\theta})}{\partial \hat{\theta}}\dot{\hat{\theta}}+ \tilde{\theta}^\top\Gamma^{-1}\dot{\hat{\theta}} \nonumber\\
&+ \frac{\partial \omega(\rho)}{\partial \rho} \dot{\rho}\big( r V(x,\hat{\theta})^{1/r} +\varepsilon \big).\label{eqn:proof_main_thm_Qdot_3nd_eq}
\end{align}
Also, utilizing the update law \eqref{eq:update.rho} in the right-hand side of \eqref{eqn:proof_main_thm_Qdot_3nd_eq} results in the cancellation of the third and fifth terms.
Thus, we have
\begin{align*}
\dot{Q}\,\leq\ 
 -\omega(\rho)\lambda V(x,\hat{\theta})^{1/r}
-\omega(\rho)V(x,\hat{\theta})^{1/r-1} \frac{\partial V(x,\hat{\theta})}{\partial x}\Delta(x)^\top\tilde{\theta} 
+\tilde{\theta}^\top\Gamma^{-1}\dot{\hat{\theta}}
\end{align*}
Finally, by substituting the  update law \eqref{eq:update.theta} into above inequality, one can see that
Consequently, with the last two terms canceled, it follows that
\begin{equation}\label{Eq:ineq.proof.int}
    \dot{Q} \leq -\omega(\rho) \lambda V(x,\hat{\theta})^{1/r} \leq 0,
\end{equation}
which implies that $Q\big(x,\hat{\theta},\omega(\rho)\big)$ is non-increasing. Hence, the value of $\omega(\rho)$, $V(x,\hat{\theta})$, $x(t)$ and $\hat{\theta}(t)$ remain bounded. Moreover, by taking the time-integral from both sides of the inequality \eqref{Eq:ineq.proof.int}, we have
\begin{align*}
    \int_0^\infty \lambda\omega(\rho) V(x,\hat{\theta})^{1/r} \mathrm{d}t &\leq 
    Q\big(x(0),\hat{\theta}(0),\omega\big(\rho(0)\big) \big)
-\lim_{t\to\infty}Q\big(x(t),\hat{\theta}(t),\omega\big(\rho(t)\big) \big) \\
  &\leq  
    Q\big(x(0),\hat{\theta}(0),\omega\big(\rho(0)\big) \big),     
\end{align*}
where the second inequality follows from the fact that $Q(x, \hat{\theta}, \omega(\rho))$ is a non-negative function. Since \(\omega(\rho)\) is strictly positive, the inequality above implies that \({V(x,\hat{\theta})}^{1/r} \in\mathcal{L}_1\). Moreover, from the inequality \eqref{eq.clf.a}, we have 
$
    \int_0^\infty ||x||_2^{2/r}\mathrm{d}t \leq \int_0^\infty  \frac{1}{c_1^{1/r}} V(x,\hat{\theta})^{1/r} \mathrm{d}t < \infty, 
$
meaning that $||x||_2^{2/r} \in \mathcal{L}_1$.
Furthermore, we have
\begin{equation*}
    \dot{V}(x,\hat{\theta})=
    \frac{\partial V(x,\hat{\theta})}{\partial x} \Big(f(x) +\Delta(x)^\top \theta 
    +B(x) u \Big)
    +\frac{\partial V(x,\hat{\theta})}{\partial \hat{\theta}}\Gamma \omega(\rho)
    V(x,\hat{\theta})^{1/r - 1}
    \Delta(x)
    \frac{\partial V(x,\hat{\theta})}{\partial x}^\top\!\!. 
\end{equation*}
Note that for $r \le vd(\Delta,V)$, the last term on the right-hand side of the above expression remains bounded. This implies that $\dot{V}(x,\hat{\theta})$ is bounded. Consequently, $V(x,\hat{\theta})$ is uniformly continuous. Since $V(x,\hat{\theta})$ is both integrable and uniformly continuous, Barbalat's lemma ensures that $V(x,\hat{\theta})$ converges to zero as $t \to \infty$, which in turn implies that $x(t) \to 0$.
\qed

\subsection{Proof of Theorem~\ref{theorem.vanish}}\label{sec:appen:thm.vanish}
Since $V(x,\theta)$ is a smooth and positive definite function with $V(0,\theta)=0$, it has a Taylor expansion of the form
\begin{equation}\label{eqn:proof_Theorem4_V_taylor_expansion}
    V(x,\theta)=\frac{1}{2}x^\top H(\theta)x+R_V(x,\theta),
\end{equation}
where $R_V: \mathbb{R}^n\times \mathbb{R}^p \to \mathbb{R}$ is a smooth function satisfying 
\begin{equation}\label{eqn:proof_Theorem4_V_taylor_expansion_res_term}
    \lim_{x\rightarrow 0} \frac{R_V(x,\theta)}{\|x\|_2^2}=0,
\end{equation}
and $H(\theta)$ is a positive definite matrix, that satisfies $c_1 I_n\leq H(\theta) \leq c_2 I_n$, for all $\theta \in \Theta$. 
Note that, due to positivity and smoothness, the gradient of $V(x,\theta)$ vanishes at zero, and therefore, the expansion \eqref{eqn:proof_Theorem4_V_taylor_expansion} does not have a linear term. 
Furthermore, since $\Delta(x)^\top$ is twice continuously differentiable, each column of it has a first-order Taylor expansion of the form
$
    [\Delta(x)^\top]_i=F_ix+R_i(x),
$
where  $F_i\in \mathbb{R}^{n \times n}$ and $R_i:\mathbb{R}^n \to \mathbb{R}^n$ is a function satisfying
$
    \lim_{x\rightarrow 0} \frac{R_i(x)}{\|x\|_2}=0.
$
Thus, using the above expressions, 
for all $r \in \mathbb{R}_{>0}$ and $i\in \{1,\ldots,p\}$,
we have
\begin{align*}
    \limsup_{x \to 0}
    &
    \,\Big|
    V(x,\theta)^{1/r - 1}\,\frac{ \partial V(x,\theta)}{\partial x}\,[\Delta(x)^\top]_i\Big|
    \, 
    \\=& \limsup_{x \to 0}
    \Big| V(x,\theta)^{1/r-1} \frac{ \partial V(x,\theta)}{\partial x}\,(F_i x + R_i(x))\Big|
    \\=& 
    \limsup_{x \to 0}
    \Big| V(x,\theta)^{1/r} \frac{ x^\top H F_i  x}{\frac{1}{2}x^\top Hx+R_V(x,\theta)} + \eta_i(x,\theta)\Big|,
\end{align*}
where $\eta_i(x,\theta)$ corresponds the higher order terms in the resulting expansion.
Accordingly, it follows that
\begin{equation}
\label{eqn:proof_Theorem4_eq_6}
    \begin{split}
    &\!\!\!\!
    \limsup_{x \to 0}
    \Big|
    V(x,\theta)^{1/r - 1}\,\frac{ \partial V(x,\theta)}{\partial x}\,[\Delta(x)^\top]_i\Big|\, \\
    &
    \le \limsup_{x \to 0}
    \Big| 
    V(x,\theta)^{1/r}
    \, 
    \frac{ x^\top H F_i  x}{x^\top Hx}
    \, 
    \frac{x^\top Hx}{\frac{1}{2}x^\top Hx+R_V(x,\theta)} 
    \Big|.
    \!\!\!\!
\end{split}    
\end{equation} 
From the positive-definiteness of $H$, one can easily see that
\begin{equation*}
    \frac{1}{\lambda_{\max}(H)}
    \frac{|R_V(x,\theta)|}{\|x\|_2^2}
    \le 
    \Big| 
    \frac{R_V(x,\theta)}{x^\top Hx} 
    \Big|
    \le 
    \frac{1}{\lambda_{\min}(H)}
    \frac{|R_V(x,\theta)|}{\|x\|_2^2}.
\end{equation*}
Consequently, \eqref{eqn:proof_Theorem4_V_taylor_expansion_res_term} implies that 
\begin{equation}
\label{eqn:proof_Theorem4_eq_7}
\lim_{x \to 0}
    \left| 
    \frac{x^\top Hx}{\frac{1}{2}x^\top Hx+R_V(x,\theta)} 
    \right|
    =2.
\end{equation}
Also, from the definition of matrix operator norm, we have
\begin{equation}\label{eqn:proof_Theorem4_eq_8}
\limsup_{x \to 0}
    \left| 
    \frac{x^\top H F_i  x}{x^\top Hx} 
    \right|
    \le
    \|H^{\frac12} F_i H^{-\frac12}\|.
\end{equation}
Subsequently, from 
\eqref{eqn:proof_Theorem4_eq_6},
\eqref{eqn:proof_Theorem4_eq_7},
and
\eqref{eqn:proof_Theorem4_eq_8},
it follows that 
\begin{align*}
    \ \ \ \ \!\!\!\!\!\!\!\!
    \limsup_{x \to 0}
    \Big|
    V(x,\theta)^{1/r - 1}\,\frac{ \partial V(x,\theta)}{\partial x}\,[\Delta(x)^\top]_i\Big|\, 
    \le 
    \,
    2 
    \|H^{\frac12} F_i H^{-\frac12}\|
    \,
    \limsup_{x \to 0}
    V(x,\theta)^{1/r}
    =
    0.
\end{align*}
As a result, one can see that    
\begin{align*}
    \lim_{x \to 0}&
\left\|
V(x,\theta)^{1/r - 1}\,\frac{ \partial V(x,\theta)}{\partial x}\,\Delta(x)^\top\, 
\right\|_{\infty}= \lim_{x \to 0} \max_i \left|   V(x,\theta)^{1/r - 1}\,\frac{ \partial V(x,\theta)}{\partial x}\,[\Delta(x)^\top]_i  \right| =0,
\end{align*}
for any  $r \in \mathbb{R}_{>0}$.
Therefore, following the definition of the vanishing degree, we have 
\begin{equation*}
\begin{split}
vd(\Delta,V) \,=\,
\sup\Bigl\{r \in \mathbb{R}_{>0} \;\Big|\;  
    \lim_{x \to 0}
      \big\|
        V(x,\theta)^{1/r - 1}\frac{ \partial V(x,\theta)}{\partial x}\,\Delta(x)^\top 
      \big\|_{\infty}
    < \infty, \forall\theta \in \Theta \Big\}= \infty,
\end{split}
\end{equation*} 
as claimed.
\qed

\vspace{-0.4cm}

\subsection{Proof of Theorem~\ref{thm:momentum}} \label{sec:appen:thm.momentum} 

Consider the non-negative function
\begin{align*}
    Q\big(x,\hat{a},\hat{\theta}\big)
    =  \,
    \omega(\rho) \big( r V(x,\hat{\theta})^{1/r} + \varepsilon \big)
    +
    \frac{1}{2}(\hat{a}-\theta)^\top\Gamma^{-1}(\hat{a}-\theta)
   + 
    \frac{1}{2}(\hat{\theta}-\hat{a})^\top \Gamma^{-1}(\hat{\theta}-\hat{a}) .
\end{align*}
Taking the time derivative of the above function along \eqref{Eq.dynamic} gives 
\begin{align*}
\dot{Q} &= \omega(\rho) V(x,\hat{\theta})^{1/r-1} 
           \frac{\partial V(x,\hat{\theta})}{\partial x} 
           \big(f(x) + \Delta(x)^\top \theta + B(x) u \big) \\
&\quad + \omega(\rho) V(x,\hat{\theta})^{1/r-1} 
           \frac{\partial V(x,\hat{\theta})}{\partial \theta} \dot{\hat{\theta}}  + \frac{\partial \omega(\rho)}{\partial \rho} \dot{\rho} 
           \big( r V(x,\hat{\theta})^{1/r} \\&\quad + \varepsilon \big) 
+ (\hat{a} - \theta)^\top \Gamma^{-1} \dot{\hat{a}} 
           + (\hat{\theta} - \hat{a})^\top \Gamma^{-1} (\dot{\hat{\theta}} - \dot{\hat{a}}).
\end{align*}
The above equation can be written as 
\small
\begin{align*}
\dot{Q} &= \omega(\rho) V(x,\hat{\theta})^{1/r-1} 
           \frac{\partial V(x,\hat{\theta})}{\partial x} 
           \big(f(x) + \Delta(x)^\top \hat{\theta} + B(x) u \big) \\
&\quad + \omega(\rho) V(x,\hat{\theta})^{1/r-1} 
           \frac{\partial V(x,\hat{\theta})}{\partial \theta} \dot{\hat{\theta}}  + \frac{\partial \omega(\rho)}{\partial \rho} \dot{\rho} 
           \big( r V(x,\hat{\theta})^{1/r} + \varepsilon \big) \\
&\quad - \omega(\rho) V(x,\hat{\theta})^{1/r-1} 
           \frac{\partial V(x,\hat{\theta})}{\partial x} \Delta(x)^\top \tilde{\theta}  + (2\hat{a} - \theta - \hat{\theta})^\top \Gamma^{-1} \dot{\hat{a}} 
           + (\hat{\theta} - \hat{a})^\top \Gamma^{-1} \dot{\hat{\theta}}.
\end{align*}
\normalsize
By substituting the control law \eqref{eq.C} and the update law \eqref{eq:update.a.col}, we have
\begin{align*}
    \dot{Q}
    \leq -\lambda \omega(\rho) V(x,\hat{\theta})^{1/r}
    +
    2\omega(\rho)V(x,\hat{\theta})^{1/r-1}\frac{\partial V(x,\theta)}{\partial x}  \Delta(x)^\top(\hat{a}-\hat{\theta})
    +(\hat{\theta}-\hat{a})^\top\Gamma^{-1}\dot{\hat{\theta}}.
\end{align*}
By setting $\dot{\hat{\theta}}$ according to the update law \eqref{eq:update.theta.col}, we have
\begin{align}
    \dot{Q}\leq -&\frac{\lambda}{2} \omega(\rho) V(x,\hat{\theta})^{1/r}
    -\lambda\omega(\rho) V(x,\hat{\theta})^{1/r-1} \Bigg( \frac{1}{2}V(x,\hat{\theta})
    \nonumber\\&\;\;\;\;\ \
    - \frac{2}{\lambda}\frac{\partial V(x,\hat{\theta})}{\partial x}  \Delta(x)^\top(\hat{a}-\hat{\theta}) 
    +\frac{4}{\lambda^2}(\hat{\theta}-\hat{a})^\top\Delta(x)P(\hat{\theta})\Delta(x)^\top(\hat{\theta}-\hat{a}) \!\Bigg).
    \label{eq.col.proof.Q}
\end{align}
Note that, in Theorem~\ref{thm:momentum}, we have
$V(x,\hat{\theta})=\frac{1}{2}x^\top P(\hat{\theta}) x$ and $\frac{\partial V(x,\hat{\theta}) }{\partial x}=x^\top P(\hat{\theta})$.
Substituting these terms into \eqref{eq.col.proof.Q}
 makes the expression inside the parentheses a perfect square, as follows
\begin{align*}
    \dot{Q}\!\leq \!-\frac{\lambda}{2} \omega(\rho) V(x,\hat{\theta})^{1/r} 
    -\frac{\lambda}{4}\omega(\rho) V(x,\hat{\theta})^{1/r-1}\times\big( x\!- \frac{4}{\lambda}\!\Delta(x)^\top (\hat{a}-\hat{\theta})\big)\!^\top\!
    P(\hat{\theta})\big( x\!- \! \frac{4}{\lambda}\Delta(x)^\top(\hat{a}-\hat{\theta})\big)\!\leq\! 0.
\end{align*}
Therefore, we conclude that the function $Q\big(x,\hat{a},\hat{\theta},\omega\big)$ is non-increasing. Hence, the signals $x(t)$, $\hat{\theta}(t)$ and $\hat{a}(t)$ are bounded. 
Furthermore, by taking the time integral
from both sides of the above inequality, we find that
$\int_0^\infty\lambda\omega(\rho) V(x,\hat{\theta})^{1/r} \mathrm{d}t 
    \leq2Q\Big(x(0),\hat{\theta}(0),\omega\big(\rho(0)\big) \Big)
    < \infty,$
which implies that $V(x,\hat{\theta})^{1/r} \in \mathcal{L}_1$. Hence, from the inequality \eqref{eq.clf.a}, we conclude that $\|x\|_2^{2/r} \in \mathcal{L}_1$. Moreover, we have
\small
\begin{align*}
    \dot{V}(x,\hat{\theta})&=\frac{\partial V(x,\hat{\theta})}{\partial x} \Big(f(x) +\Delta(x)^\top 
    \theta 
    +
    B(x) u \Big)
    +\frac{\partial V(x,\hat{\theta})}{\partial \hat{\theta}}\dot{\hat{\theta}}. 
\end{align*}
\normalsize
The assumptions stated in Theorem~\ref{thm:momentum} ensure that the update law $\dot{\hat{\theta}}$ remains bounded. As a result, the right-hand side of the above expression is bounded, which implies $V(x,\hat{\theta})$ is uniformly continuous. Hence, by Barbalat's lemma, $V(x,\hat{\theta})$ and, consequently, $x(t)$ converge to zero as $t \to \infty$.
\section{Numerical Results}\label{sec:sim}
In this section, we demonstrate the performance improvements achieved by the proposed parameter estimation laws using three numerical examples, each representing a different class of dynamics and uncertainty structure. The first example considers a system with a linear nominal term and a nonlinear matched uncertainty; the second addresses systems with unmatched uncertainty in parametric strict-feedback form; and the third investigates a robot manipulator with flexible joints.

\begin{figure}[t]
    \centering
    \begin{subfigure}{0.32\columnwidth}
        \centering
        \includegraphics[width=\textwidth]{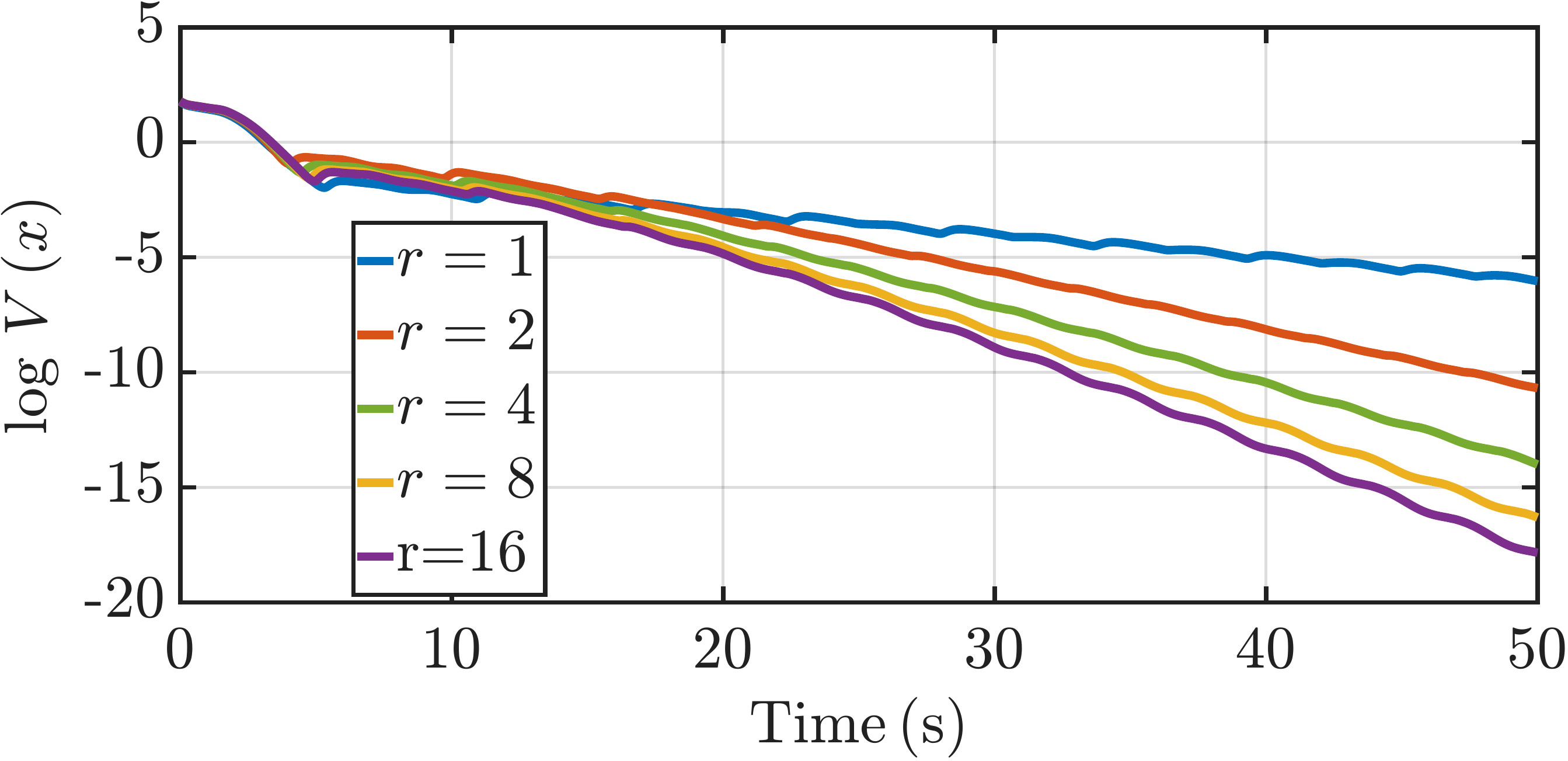}
        \caption{Evolution of $\log V(x,\hat{\theta})$}
        \label{fig:Slope_a}
    \end{subfigure}
    \hfill
    \begin{subfigure}{0.32\columnwidth}
        \centering
        \includegraphics[width=\textwidth]{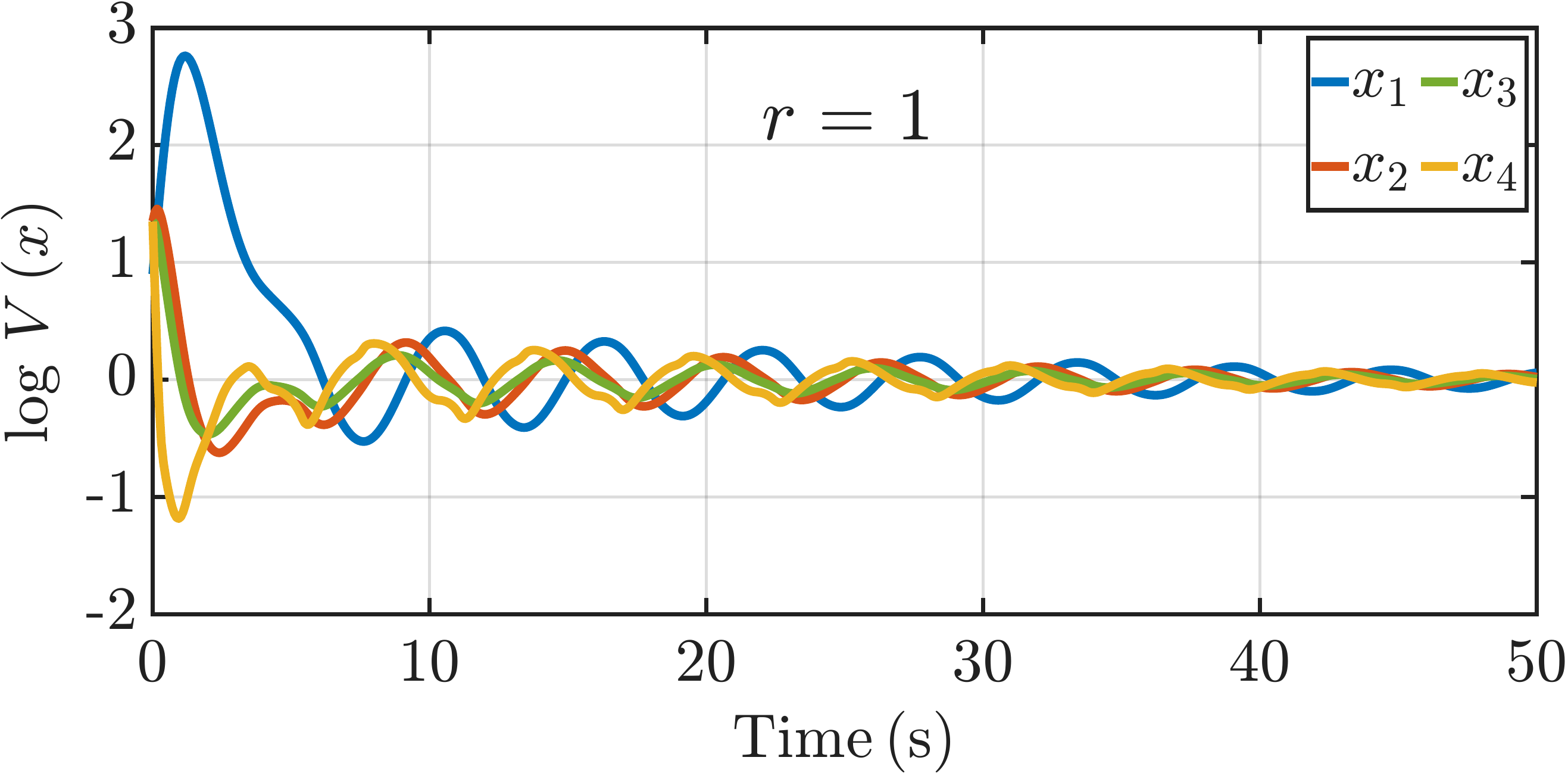}
        \caption{Evolution of state for $r=1$}
        \label{fig:Slope_b}
    \end{subfigure}
    \hfill
    \begin{subfigure}{0.32\columnwidth}
        \centering
        \includegraphics[width=\textwidth]{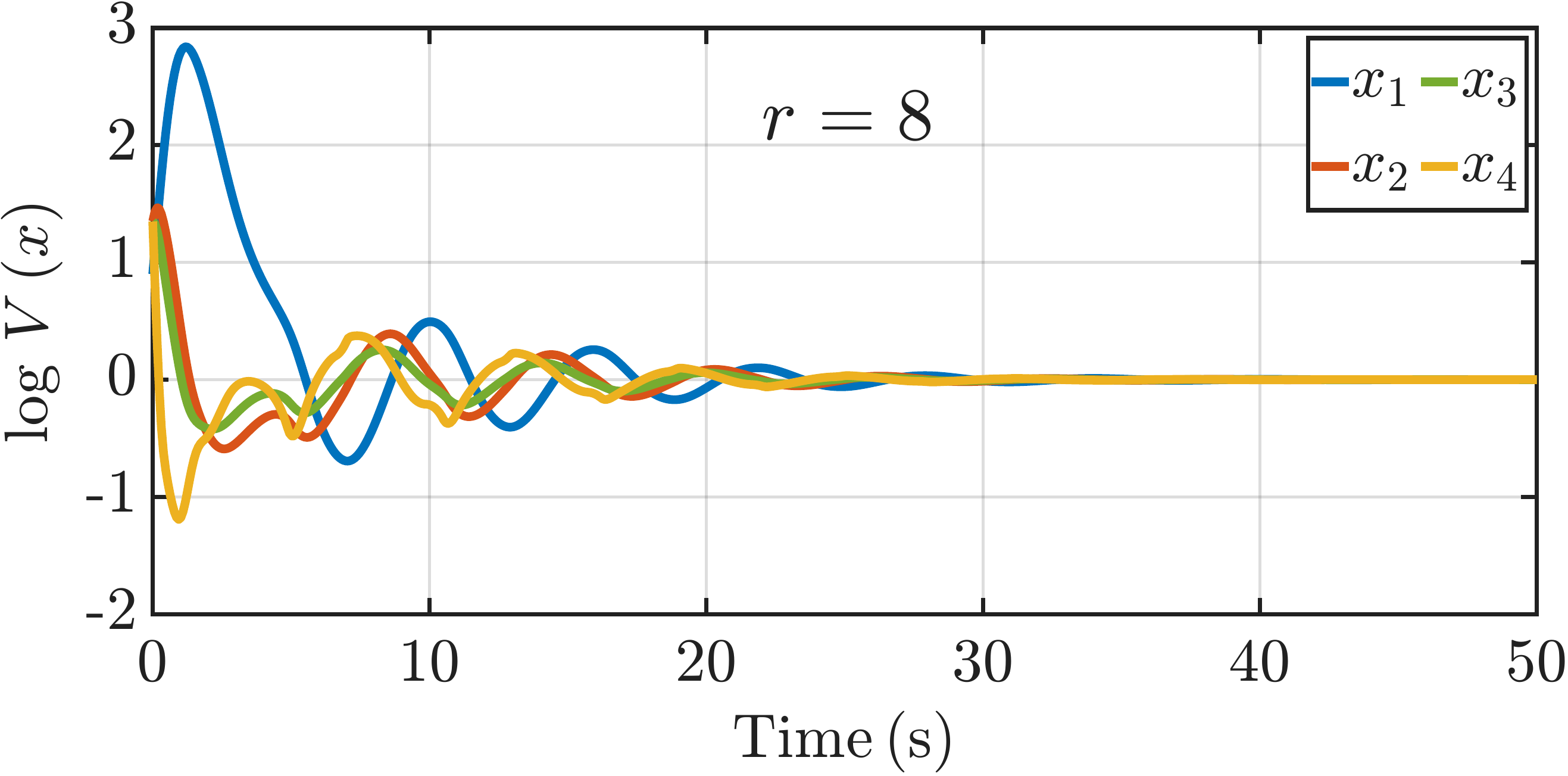}
        \caption{Evolution of state for $r=8$}
        \label{fig:Slope_c}
    \end{subfigure}

    \caption{(a) Plot of the log Lyapunov function versus time for Example~\ref{examples.matched}, where the control law \eqref{eq.C} and the update laws \eqref{eq:update.theta} are employed for different values of \( r \).  
    (b)--(c) Closed-loop state variables in Example~\ref{examples.matched}, where the update law \eqref{eq:update.theta} is employed for \( r=1 \) and \( r=8 \), respectively.}
\end{figure}
\begin{figure}[t]
    \centering
    \begin{subfigure}{0.32\columnwidth}
        \centering
        \includegraphics[width=\textwidth]{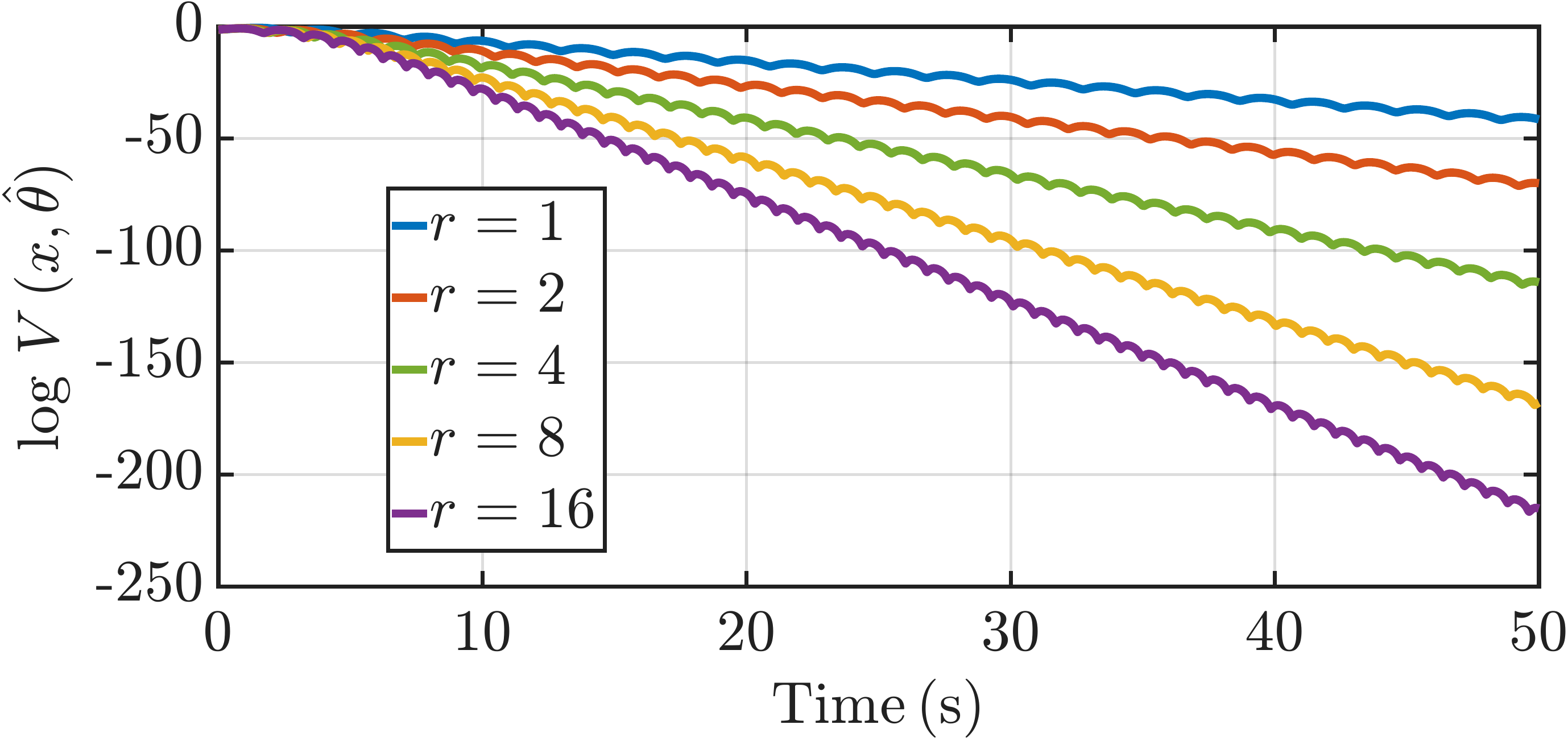}
        \caption{Evolution of $\log V(x,\hat{\theta})$}
        \label{fig:ex2_Slope_a}
    \end{subfigure}
    \hfill
    \begin{subfigure}{0.32\columnwidth}
        \centering
        \includegraphics[width=\textwidth]{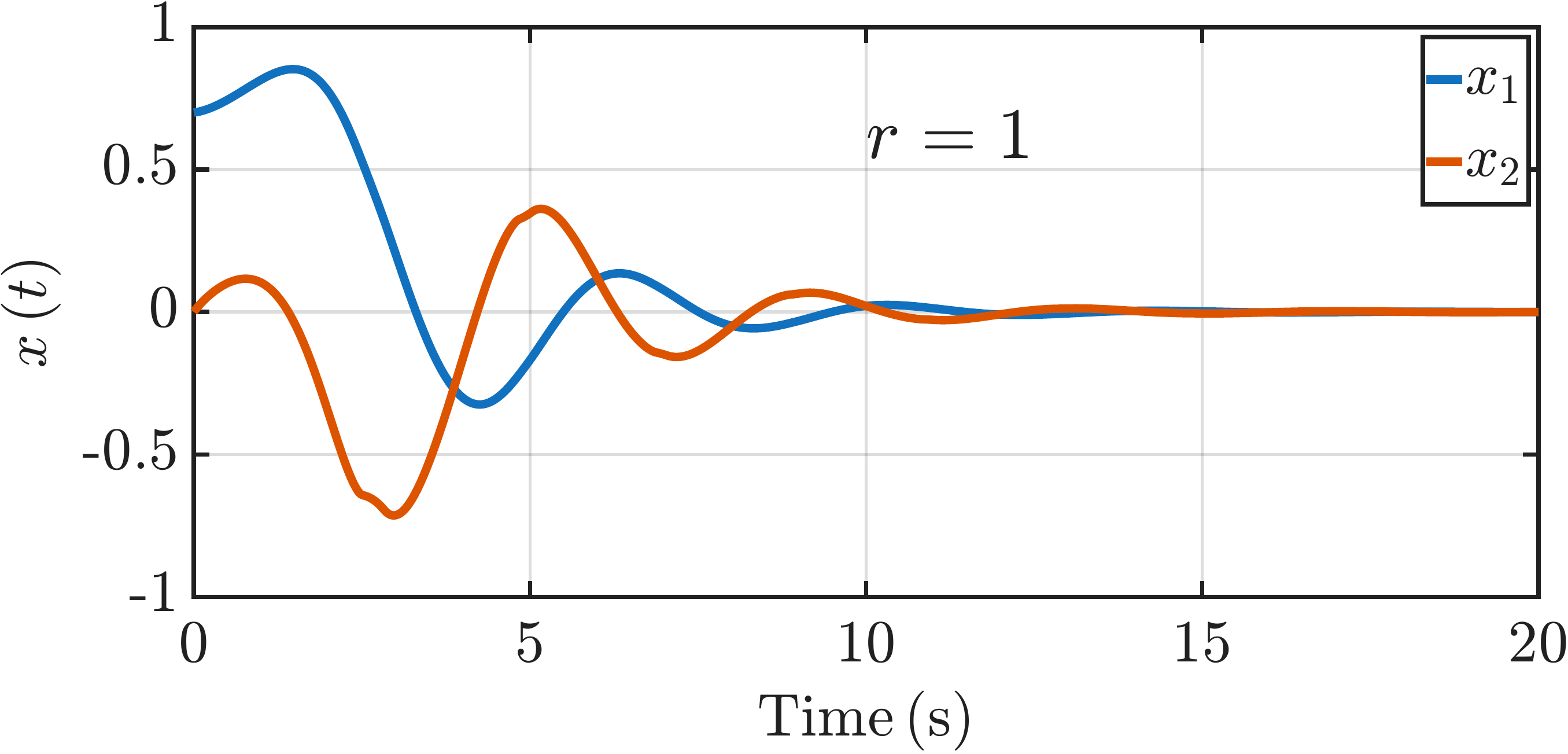}
        \caption{Evolution of state for $r=1$}
        \label{fig:ex2_Slope_b}
    \end{subfigure}
    \hfill
    \begin{subfigure}{0.32\columnwidth}
        \centering
        \includegraphics[width=\textwidth]{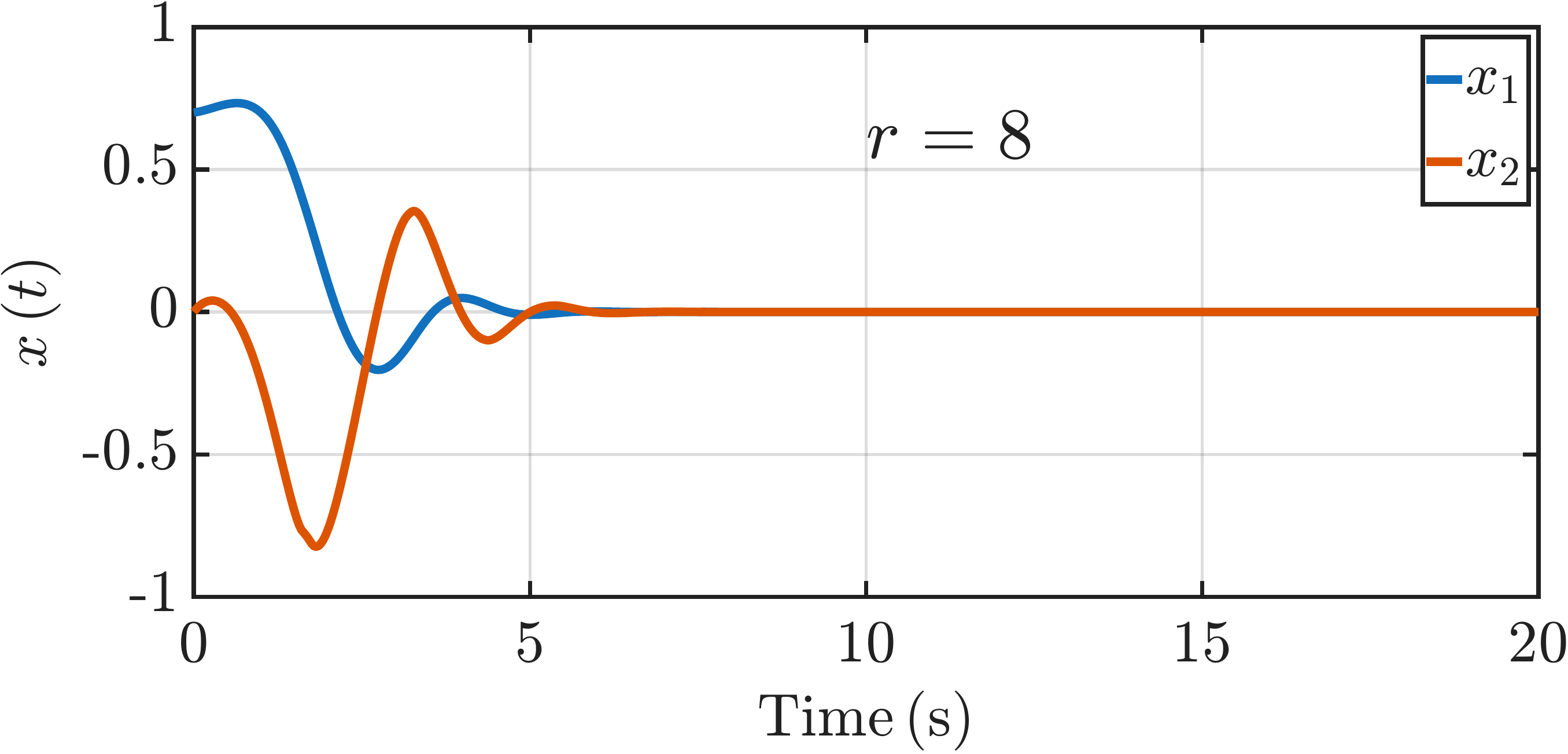}
        \caption{Evolution of state for $r=8$}
        \label{fig:ex2_Slope_c}
    \end{subfigure}

    \caption{(a) Plot of the log Lyapunov function versus time for Example~2, where the control law \eqref{eq.C} and the update laws \eqref{eq:update.theta} are employed for different values of \( r \).  
    (b)--(c) Closed-loop state variables in Example~\ref{examples.unmached}, where the update laws \eqref{eq:update.theta} and \eqref{eq:update.rho} are employed for \( r=1 \) and \( r=8 \), respectively.}
    \label{fig:combined_ex2}
\end{figure}
\begin{figure}[t]
    \centering
    \begin{subfigure}{0.32\columnwidth}
        \centering
        \includegraphics[width=\textwidth]{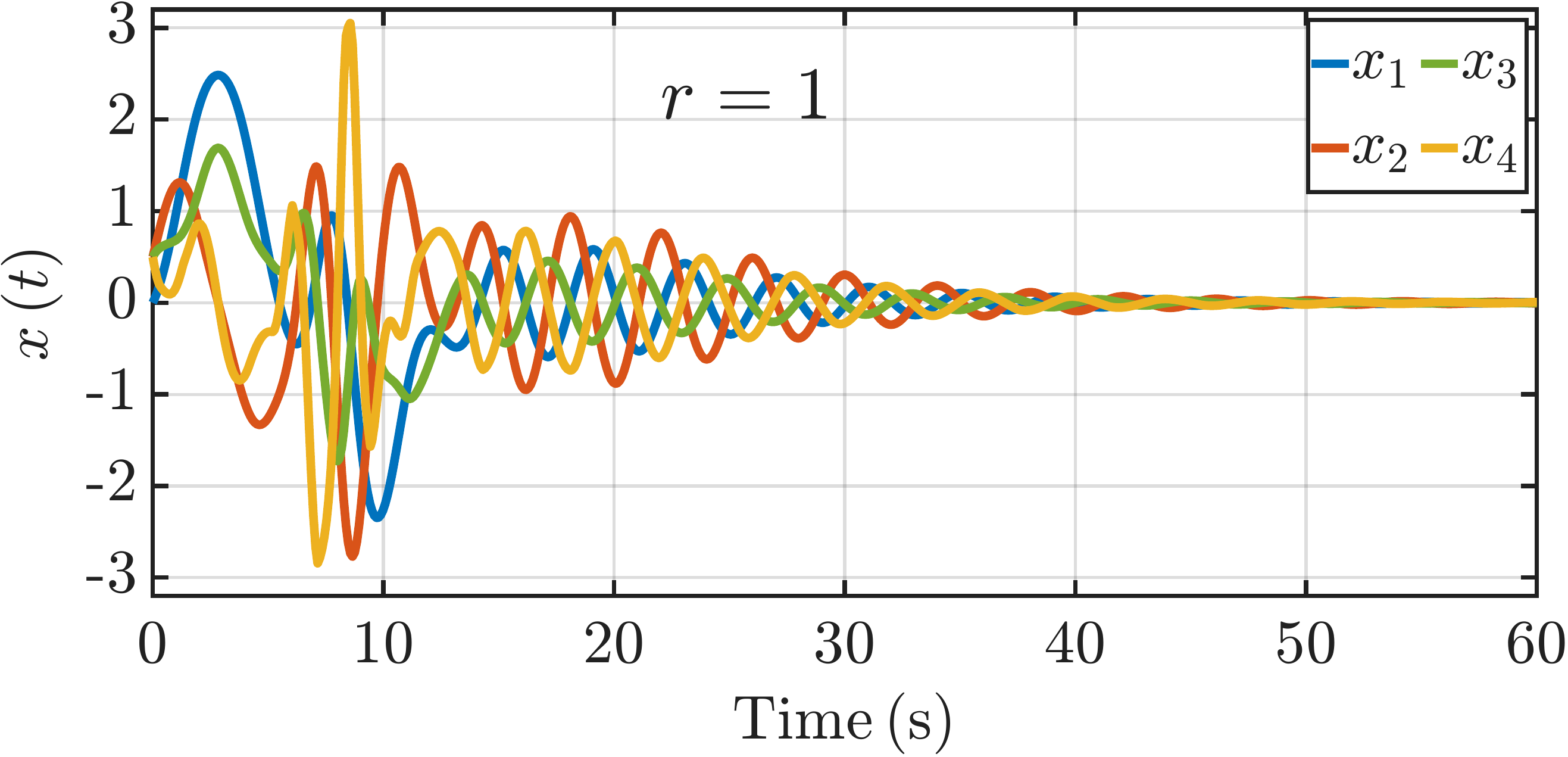}
        \caption{Evolution of state for $r=1$}
        \label{fig:ex3_n1}
    \end{subfigure}
    \hfill
    \begin{subfigure}{0.32\columnwidth}
        \centering
        \includegraphics[width=\textwidth]{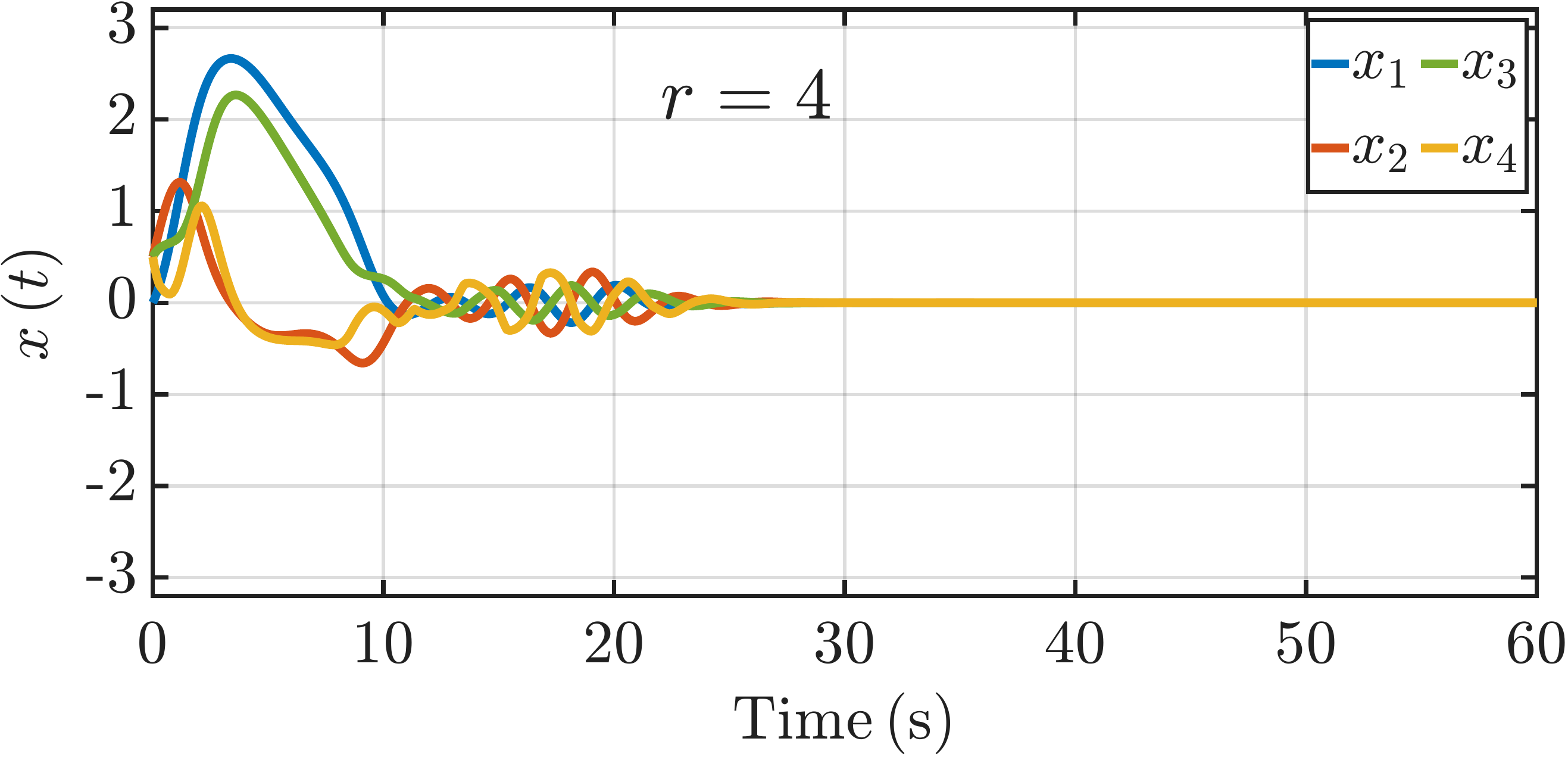}
        \caption{Evolution of state for $r=4$}
        \label{fig:ex3_n4}
    \end{subfigure}
    \hfill
    \begin{subfigure}{0.32\columnwidth}
        \centering
        \includegraphics[width=\textwidth]{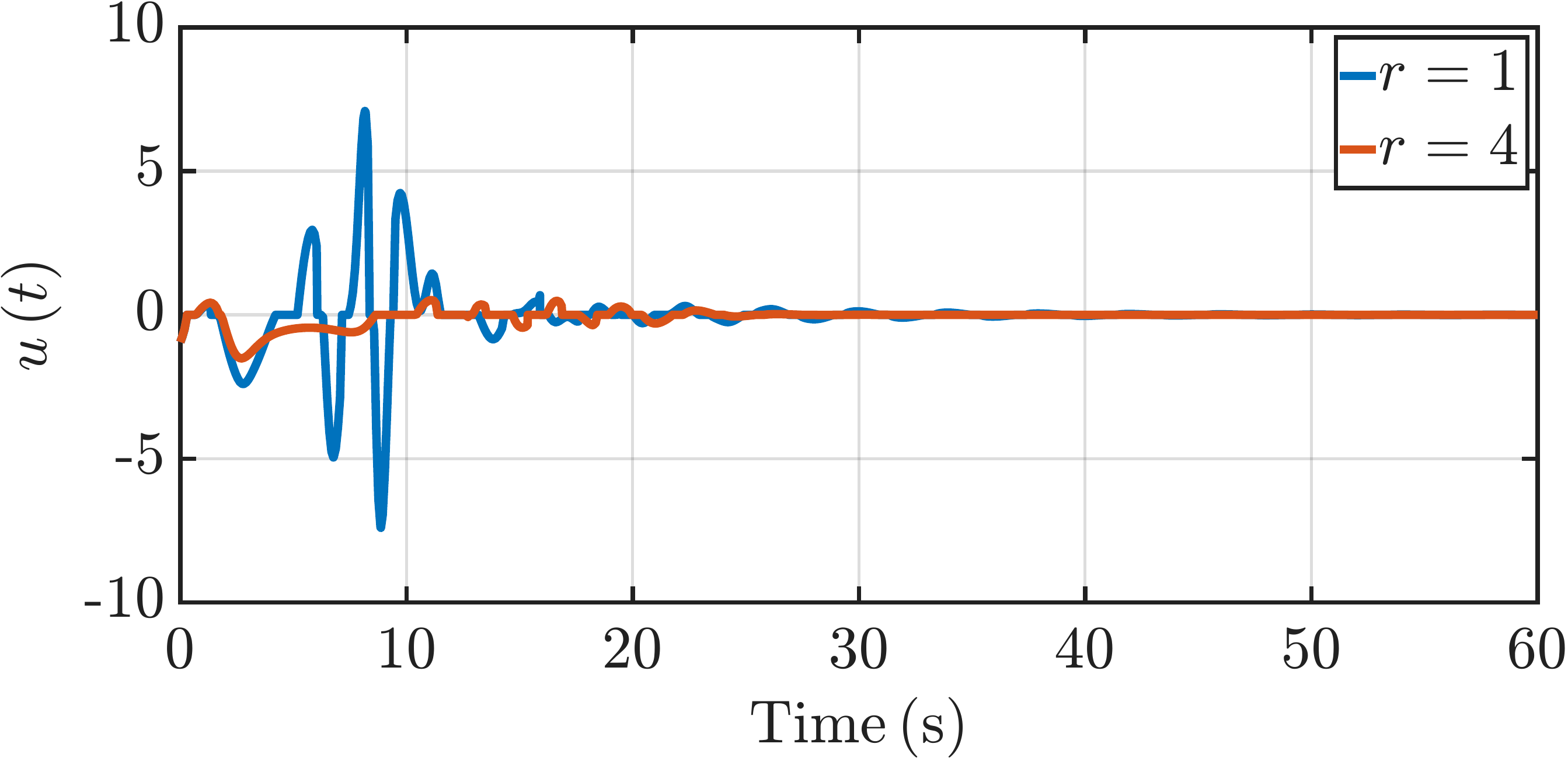}
        \caption{$u(t)$ for $r = 1$ and $4$}
        \label{fig:ex3_u}
    \end{subfigure}

    \caption{(a)--(b) Closed-loop state variables in Example~\ref{examples.robot}, using the update laws from Theorem~\ref{theorem_main} with \( r=1 \) and \( r=4 \), respectively.  
    (c) Control signal \( u(t) \) in Example~\ref{examples.robot} using the update laws from Theorem~\ref{theorem_main} with \( r=1 \) and \( r=4 \).}
    \label{fig:combined_ex3}
\end{figure}

\begin{example}[Matched uncertainty] \label{examples.matched}
\normalfont
 Consider the dynamical system
\begin{equation}\label{Eq:sim.ex.sys}
    \dot{x} = Ax + B\big(u + \varphi(x)^\top \theta\big).
\end{equation}
Note that the uncertainty term \(B\,\varphi(x)^\top \theta\) satisfies the matching condition described in Remark~\ref{remark.match}.
By applying Finsler's lemma \cite{UHLIG1979219} to \eqref{eq.clf}, we obtain that any quadratic function of the form \( V(x) = x^\top P^{-1} x \) serves as a CLF for system \eqref{Eq:sim.ex.sys} if and only if the positive definite matrix $P$ satisfies the linear matrix inequality
\begin{equation}\label{eq:lmi}
    B^\top_\perp (AP + PA^\top + \lambda P) B_\perp \leq 0,
\end{equation}
where \( B_\perp \) is the annihilator matrix of \( B \), i.e., \( B^\top_\perp B = 0 \).
For our numerical demonstration, we set 
$A$, $B$, and $\varphi(x)$ respectively as  
\small 
\begin{equation*}
    A
    =
    \!
    \begin{bmatrix}
        0 & 1 & 1 & 0  \\[2pt]
        0 & -1 & 1 &1 \\[2pt]
        0 & 0 & -1 & 1 \\[2pt]
        0 & 0   &  0 &0
    \end{bmatrix}, 
    \ 
    B = 
    \!
    \begin{bmatrix}
    0 \\[2pt]
    0 \\[2pt]
    0 \\[2pt]
    1
    \end{bmatrix}\!,
    \  
    \varphi(x)= 
    \!
    \begin{bmatrix}
         x_1^2  \\[2pt]
        \arctan\left(x_2\right) \\[2pt]
        x_3 \\[2pt]
        \arctan(x_4)
    \end{bmatrix}\!.
\end{equation*}
\normalsize
Following Theorem~\ref{theorem.vanish}, we have that the
vanishing degree \( vd(\Delta,V) \) is infinite,
implying that the parameter \( r \) in Theorem~\ref{theorem_main} can be chosen arbitrarily large.
In the rest of this example, we set the true system parameter vector to 
$
\theta = \begin{bmatrix} 0.5,\, 1,\, 1,\, 1 \end{bmatrix}^\top.
$
The system state and the parameter estimate are initialized as  
$
x(0) = \begin{bmatrix} 1,\, 1.5,\, 1.5,\, 1.5 \end{bmatrix}^\top,$ and $
\hat{\theta}(0) = \begin{bmatrix} 0,\, 0,\, 0,\, 0 \end{bmatrix}^\top.
$
The adaptation gain is set to \(\Gamma = I_{4}\).

Figure~\ref{fig:Slope_a} illustrates the log of \( V(x) \) as a function of time. Each curve is obtained by applying the controller \eqref{eq.theore.C} and the parameter update law \eqref{eq:update.theta} for a fixed value of \( r \), namely 
$r\in\{1,2,4,8,16\}$. 
As we can observe, increasing \( r \) accelerates the convergence of \( V(x) \). Figure~\ref{fig:Slope_b} and Figure~\ref{fig:Slope_c} present the system state trajectories for \( r = 1 \) and \( r = 8 \), respectively. It can be seen that the states converge to zero significantly faster for \( r = 8 \) compared to \( r = 1 \).
\end{example}

\begin{example}[Unmatched uncertainty] \label{examples.unmached}\normalfont
  Consider the following system with the parametric strict-feedback form as    
\begin{equation}
\begin{split}
    \dot{x}_1 &= x_2 + \theta^\top \varphi_1(x_1), \\
    \dot{x}_2 &= \theta^\top \varphi_2(x_1, x_2) + u,
\end{split}  
\end{equation}
where \( x = [x_1, x_2]^\top \) is the system state, \( u \) is the control input, \( \theta \in \mathbb{R}^p \) is an unknown parameter vector, and \( \varphi_1(x_1) \), \( \varphi_2(x_1, x_2) \) are known functions.  Using the back-stepping technique \cite{Krstic}, we construct the following CLF
\begin{equation*}
    V(x, \theta) = \frac{1}{2}x_1^2 + \frac{1}{2} \big(x_2 + \theta^\top \varphi_1(x_1) + \lambda x_1 \big)^2,
\end{equation*}
for a given $\lambda>0$.
For our numerical simulations, we assume
\begin{equation*}
\varphi_1(x_1) = 
\begin{bmatrix}
\sin(x_1) \\[4pt]
\arctan(x_1) \\[4pt]
0 \\[4pt]
0
\end{bmatrix}, \quad
\varphi_2(x_1, x_2) = 
\begin{bmatrix}
0 \\[4pt]
0 \\[4pt]
\sin(x_1) \\[4pt]
\arctan(x_2)
\end{bmatrix}.
\end{equation*}
Note that in this example, $V(x,\theta)$ is a function of the system parameters. 
Moreover, according to Theorem~\ref{theorem.vanish}, one can see that the vanishing degree $vd(\Delta,V)$ is infinite.

We set the design parameters in our estimation law as
\[
\Gamma = 0.2 I_{4}, \quad \omega(\rho) = \frac{\pi}{2} + \arctan(0.5\,\rho), \quad \varepsilon = 0.5.
\]
Figure~\ref{fig:ex2_Slope_a} presents the logarithm of \(V(x, \theta)\) over time. 
Similar to the previous example, each curve corresponds to a fixed \(r\), for $r\in\{1,2,4,8,16\}$, and is obtained 
using the controller in~\eqref{eq.theore.C} together with the parameter estimation law 
from Theorem~\ref{theorem_main}. Furthermore, Figure~\ref{fig:ex2_Slope_b} and Figure~\ref{fig:ex2_Slope_c} present the system state trajectories for \( r = 1 \) and \( r = 8 \), respectively. We can observe that increasing $r$ 
accelerates convergence.   
\end{example}




%
\begin{example}[Manipulator with flexible joints]\label{examples.robot}
\normalfont
The dynamics of a single link manipulator with flexible joints  can be described as \cite{khalil2002nonlinear}
\begin{equation}\label{eqn:ex_manipulator}
\begin{split}  
\dot{x}_1 &= x_2, \\
\dot{x}_2 &= \theta_1 \sin x_1 + \theta_2 (x_3 - x_1), \\
\dot{x}_3 &= x_4, \\
\dot{x}_4 &=\theta_3 (x_1 - x_3) +\theta_4 x_4 + u,
\end{split}
\end{equation}
where $x_1$ and $x_3$ are angular positions, $x_2$ and $x_4$ are angular velocities, and 
$\theta = [\theta_1 \;\; \theta_2 \;\; \theta_3 \;\; \theta_4]^\top$ is an unknown parameter vector, with $\theta_2 > 0$. 
We note that system \eqref{eqn:ex_manipulator} is feedback linearizable, and therefore, a CLF can be obtained using feedback linearization techniques~ \cite{freeman2008robust}. 
In particular, one can verify that  the function
\begin{equation*}
    V(x,\theta) = \Psi(x,\theta)^\top\, P^{-1}\, \Psi(x,\theta),
\end{equation*}
where  
\begin{equation*}
\Psi(x,\theta) =
\begin{bmatrix}
x_1 \\[4pt]
x_2 \\[4pt]
\theta_{1} \sin x_{1} + \theta_{2} (x_{3} - x_{1}) \\[4pt]
\theta_{1} x_{2}\, \cos x_{1}   + \theta_{2} (x_{4} - x_{2})
\end{bmatrix},
\end{equation*}
and \(P\) is a positive-definite matrix satisfying the linear matrix inequality~\eqref{eq:lmi} for  
\begin{equation*}
    A=\begin{bmatrix}
        0 & 1 & 0 & 0  \\
        0 & 0 & 1 & 0 \\
        0 & 0 & 0 & 1 \\
        0 & 0 & 0 & 0
    \end{bmatrix}, 
    \quad 
    B= \begin{bmatrix}
        0 \\[2pt] 0 \\[2pt] 0 \\[2pt] 1
    \end{bmatrix},
\end{equation*}
is a CLF for system \eqref{eqn:ex_manipulator}.
According to Theorem~\ref{theorem.vanish}, the vanishing degree \( vd(\Delta,V) \) associated with system \eqref{eqn:ex_manipulator} is infinite. Hence, the parameter \( r \) in Theorem~\ref{theorem_main} can be chosen arbitrarily large.

Figure~\ref{fig:ex3_n1} and Figure~\ref{fig:ex3_n4} show the system states when the parameter estimation law in Theorem~\ref{theorem_main} are employed, for  $r=1$ and $r=4$, together with the control \eqref{eq.theore.C}. 
It can be observed that the system states converge significantly faster to zero for $r=4$ compared to 
$r=1$. 
Figure~\ref{fig:ex3_u} shows the input signal $u(t)$ for the mentioned scenarios. 
One can see that the control effort for $r=4$ has a smaller magnitude and fewer oscillations.
\end{example}

\section{Conclusions And Future Directions}\label{sec:dis}
In this paper, we proposed a new family of parameter estimation laws and their higher-order extensions for adaptive systems. We demonstrated that these estimation laws can significantly accelerate the convergence of the system states by promoting signal sparsity in the time domain, i.e., by penalizing prolonged signal duration and slow decay. The developed approach is particularly advantageous for adaptive stabilization and regulation problems in which persistent excitation is absent and the primary objective is to accelerate state convergence, rather than to ensure convergence of the parameter estimates to their true values. Its benefits are most pronounced for exponentially stabilizable systems whose regressor vanishes at the origin, for which the proposed laws guarantee $\|x(t)\|_2^{2/r}\in L_1$ for arbitrarily large finite $r$, without requiring high adaptation gains or prior knowledge of the true parameter values. The method can be integrated with any certainty-equivalence CLF-based controller and applies to systems with either matched or unmatched parametric uncertainties.

Future work could extend these results to accelerate convergence in adaptive safety and control barrier functions~ \cite{ames,Brett.2}, fault compensation in dynamical systems~ \cite{Bov}, adaptive observer design~ \cite{Krstic,BOVEIRI2024100938}, and adaptive control for systems with time-varying parameters~ \cite{chen2021adaptive}.

\printbibliography

@article{lopez2019contraction,
  author       = {Lopez, Brett T. and Slotine, Jean-Jacques E.},
  title        = {Contraction Metrics in Adaptive Nonlinear Control},
  year         = {2019},
  eprint       = {1912.13138},
  eprinttype   = {arXiv},
}

@article{BOVEIRI2024100938,
  title = {Output feedback and adaptive output feedback control design for input-affine nonlinear systems based on Riemannian metrics},
  journal = {Eur. J. Control},
  volume = {75},
  pages = {100938},
  year = {2024},
  author = {Mohammad Boveiri and Mohammad Saleh Tavazoei},
}

@book{khalil2002nonlinear,
  author = {Khalil, Hassan K.},
  title = {Nonlinear systems},
  year = {2002},
  address = {Englewood Cliffs, NJ, USA},
  publisher = {Prentice-Hall},
}

@article{chen2021adaptive,
  title = {Adaptive control for systems with time‑varying parameters},
  author = {Chen, Kaiwen and Astolfi, Alessandro},
  journal = {IEEE Trans. Autom. Control},
  volume = {66},
  number = {5},
  pages = {1986--2001},
  year = {2021},
}

@book{ortega1998passivity,
  author = {Ortega, Romeo and Loría, Antonio and Nicklasso, Per Johan n and others},
  title = {Passivity-based Control of Euler-Lagrange Systems: Mechanical, Electrical and Electromechanical Applications},
  publisher = {Springer},
  year = {1998},
}

@ARTICLE{Power.1,
  author = {Donald A. Pierre},
  journal = {IEEE Trans. Power Syst.},
  title = {A perspective on adaptive control of power systems},
  year = {1987},
  volume = {2},
  number = {2},
  pages = {387--395},
}

@article{WANG2020104704,
  title = {Fixed-time control design for nonlinear uncertain systems via adaptive method},
  journal = {Syst. Control Lett.},
  volume = {140},
  pages = {104704},
  year = {2020},
  author = {Wang, Fang and Lai, Guanyu},
}

@ARTICLE{Meng-fix,
  author = {Meng, Qingtan and Ma, Qian and Shi, Yang},
  journal = {IEEE Trans. Autom. Control},
  title = {Adaptive fixed-time stabilization for a class of uncertain nonlinear systems},
  year = {2023},
  volume = {68},
  number = {11},
  pages = {6929--6936},
}

@misc{wensing2018beyond,
  title = {Beyond convexity -- contraction and global convergence of gradient descent},
  author = {Wensing, Patrick M. and Slotine, Jean-Jacques E.},
  year = {2018},
  eprint = {1806.06655},
  archivePrefix = {arXiv},
}

@inproceedings{lee2018natural,
  title = {A natural adaptive control law for robot manipulators},
  author = {Lee, Taeyoon and Kwon, Jaewoon and Park, Frank C.},
  booktitle = {Proc. IEEE/RSJ Int. Conf. Intell. Robots Syst. (IROS)},
  year = {2018},
  pages = {8628--8635},
}

@book{bach2024learning,
  title = {Learning theory from first principles},
  author = {Bach, Francis},
  year = {2024},
  publisher = {The MIT Press},
  series = {Adaptive Computation and Machine Learning},
  address = {Cambridge, MA},
}

@article{nemirovski2009robust,
  title = {Robust stochastic approximation approach to stochastic programming},
  author = {Nemirovski, Arkadi and Juditsky, Anatoli and Lan, Guanghui and others},
  journal = {SIAM J. Optim.},
  volume = {19},
  number = {4},
  pages = {1574--1609},
  year = {2009},
}

@book{nesterov2004introductory,
  title = {Introductory Lectures on Convex Optimization: A Basic Course},
  author = {Nesterov, Yurii},
  year = {2004},
  publisher = {Springer},
  series = {Applied Optimization},
  volume = {87},
}

@article{narendra1987persistent,
  title = {Persistent excitation in adaptive systems},
  author = {Narendra, Kumpati S and Annaswamy, Anuradha M},
  journal = {Int. J. Control},
  volume = {45},
  number = {1},
  pages = {127--160},
  year = {1987},
  publisher = {Taylor \& Francis},
}

@article{boyd_sastry_1986,
  title = {Necessary and sufficient conditions for parameter convergence in adaptive control},
  author = {Boyd, Stephen and Sastry, Shankar},
  journal = {Automatica},
  volume = {22},
  number = {6},
  pages = {629--639},
  year = {1986},
}

@ARTICLE{Gaudio,
  author = {Gaudio, Joseph E. and Annaswamy, Anuradha M. and Bolender, Michael A. and others},
  journal = {IEEE Control Syst. Lett.},
  title = {A class of high-order tuners for adaptive systems},
  year = {2021},
  volume = {5},
  number = {2},
  pages = {391--396},
}

@article{nesterov1983method,
  title = {A method for solving the convex programming problem with convergence rate \(O(1/k^2)\)},
  author = {Nesterov, Yurii},
  journal = {Sov. Math. Dokl.},
  volume = {27},
  number = {2},
  pages = {372--376},
  year = {1983},
  publisher = {American Mathematical Society},
}

@book{pavlov2006uniform,
  title = {Uniform output regulation of nonlinear systems: A convergent dynamics approach},
  author = {Alexey Pavlov and Nathan Wouw and Henk Nijmeijer},
  year = {2006},
  publisher = {Springer Science \& Business Media},
}

@ARTICLE{Bov,
  author = {Boveiri, Mohammad and Tavazoei, Mohammad Saleh},
  journal = {IEEE Control Syst. Lett.},
  title = {Adaptive Actuator Failure Compensation on the Basis of Contraction Metrics},
  year = {2022},
  volume = {6},
  number = {},
  pages = {1376--1381},
}

@book{boyd2004convex,
  title = {Convex Optimization},
  author = {Boyd, Stephen and Vandenberghe, Lieven},
  year = {2004},
  publisher = {Cambridge University Press},
}

@article{SLOTINE.com,
  title = {Composite adaptive control of robot manipulators},
  journal = {Automatica},
  volume = {25},
  number = {4},
  pages = {509--519},
  year = {1989},
  issn = {0005-1098},
  author = {Jean-Jacques E. Slotine and Weiping Li},
}

@article{Bech.2,
  title = {Adaptive control with guaranteed transient and steady state tracking error bounds for strict feedback systems},
  journal = {Automatica},
  volume = {45},
  number = {2},
  pages = {532--538},
  year = {2009},
  author = {Charalampos P. Bechlioulis and George A. Rovithakis},
}

@ARTICLE{Bech.1,
  author = {Bechlioulis, Charalampos P. and Rovithakis, George A.},
  journal = {IEEE Trans. Autom. Control},
  title = {Robust Adaptive Control of Feedback Linearizable MIMO Nonlinear Systems With Prescribed Performance},
  year = {2008},
  volume = {53},
  number = {9},
  pages = {2090--2099},
}

@ARTICLE{Mazenc,
  author = {Mazenc, Frederic and de Queiroz, Marcio and Malisoff, Michael},
  journal = {IEEE Trans. Autom. Control},
  title = {Uniform Global Asymptotic Stability of a Class of Adaptively Controlled Nonlinear Systems},
  year = {2009},
  volume = {54},
  number = {5},
  pages = {1152--1158},
}

@book{bastin2013line,
  title = {On-line estimation and adaptive control of bioreactors},
  author = {Bastin, Georges},
  volume = {1},
  year = {2013},
  publisher = {Elsevier},
}

@article{SONG1992271,
  title = {Study on the exponential path tracking control of robot manipulators via direct adaptive methods},
  journal = {Robot. Auton. Syst.},
  volume = {9},
  number = {4},
  pages = {271--282},
  year = {1992},
  author = {Yong-Duan Song and Richard H. Middleton and Joe N. Anderson},
}

@article{jenkins,
  author = {Jenkins, Benjamin M. and Annaswamy, Anuradha M. and Lavretsky, Eugene and others},
  title = {Convergence properties of adaptive systems and the definition of exponential stability},
  journal = {SIAM J. Control Optim.},
  volume = {56},
  number = {4},
  pages = {2463--2484},
  year = {2018},
}

@ARTICLE{driving,
  author = {Li, Xuefang and Liu, Chengyuan and Chen, Boli and others},
  journal = {IEEE Trans. Intell. Transp. Syst.},
  title = {Robust adaptive learning-based path tracking control of autonomous vehicles under uncertain driving environments},
  year = {2022},
  volume = {23},
  number = {11},
  pages = {20798--20809},
}

@ARTICLE{Brett.2,
  author = {Lopez, Brett T. and Slotine, Jean-Jacques E. and How, Jonathan P.},
  journal = {IEEE Control Syst. Lett.},
  title = {Robust adaptive control barrier functions: An adaptive and data-driven approach to safety},
  year = {2021},
  volume = {5},
  number = {3},
  pages = {1031--1036},
}

@ARTICLE{flight.2,
  author = {Bodson, Marc and Groszkiewicz, Joseph E.},
  journal = {IEEE Trans. Control Syst. Technol.},
  title = {Multivariable adaptive algorithms for reconfigurable flight control},
  year = {1997},
  volume = {5},
  number = {2},
  pages = {217--229},
}

@ARTICLE{flight.1,
  author = {Dydek, Zachary T. and Annaswamy, Anuradha M. and Lavretsky, Eugene},
  journal = {IEEE Control Syst. Mag.},
  title = {Adaptive control and the NASA X-15-3 flight revisited},
  year = {2010},
  volume = {30},
  number = {3},
  pages = {32--48},
}

@article{UHLIG1979219,
  title = {A recurring theorem about pairs of quadratic forms and extensions: a survey},
  journal = {Linear Algebra Appl.},
  volume = {25},
  pages = {219--237},
  year = {1979},
  author = {Frank Uhlig},
}

@INPROCEEDINGS{taylor,
  author = {Taylor, Andrew J. and Ames, Aaron D.},
  booktitle = {Proc. Amer. Control Conf. (ACC)},
  title = {Adaptive safety with control barrier functions},
  year = {2020},
  volume = {},
  number = {},
  pages = {1399--1405},
}

@book{ioannou1996robust,
  author = {Petros A. Ioannou and Jing Sun},
  title = {Robust adaptive control},
  publisher = {Prentice Hall},
  year = {1996},
}

@book{slotine1991applied,
  author = {Jean-Jacques E. Slotine and Weiping Li},
  title = {Applied nonlinear control},
  publisher = {Prentice Hall},
  year = {1991},
  address = {Englewood Cliffs, NJ},
}

@ARTICLE{Exp,
  author = {Song, Yongduan and Zhao, Kai and Krstic, Miroslav},
  journal = {IEEE Trans. Autom. Control},
  title = {Adaptive control with exponential regulation in the absence of persistent excitation},
  year = {2017},
  volume = {62},
  number = {5},
  pages = {2589--2596},
}

@article{spong.2022,
  author = {Spong, Mark W.},
  title = {An historical perspective on the control of robotic Manipulators},
  journal = {Annu. Rev. Control Robot. Auton. Syst.},
  year = {2022},
  volume = {5},
  pages = {1--31},
}

@article{slotine1987adaptive,
  author = {Slotine, Jean-Jacques E. and Li, Weiping},
  title = {On the adaptive control of robot manipulators},
  journal = {Int. J. Robot. Res.},
  volume = {6},
  number = {3},
  pages = {49--63},
  year = {1987},
}

@ARTICLE{Sos,
  author = {Tan, Weehong and Packard, Andrew},
  journal = {IEEE Trans. Autom. Control},
  title = {Stability region analysis using polynomial and composite polynomial Lyapunov functions and sum-of-squares programming},
  year = {2008},
  volume = {53},
  number = {2},
  pages = {565--571},
}

@article{Boffi.1,
  author = {Nicholas M. Boffi, Jean-Jacques E. Slotine},
  title = { Implicit regularization and momentum algorithms in nonlinearly parameterized adaptive control and prediction},
  journal = {Neural Comput.},
  year = {2021},
  volume = {33},
  pages = {590--673},
}

@article{Lopez.universal,
  author = {Brett T. Lopez and Jean-Jacques E. Slotine},
  title = {Universal adaptive control of nonlinear systems},
  journal = {IEEE Control Syst. Lett.},
  year = {2021},
}

@book{Krstic,
  publisher = {John Willey},
  author = {Miroslav Krstic and Ioannis Kanellakopoulos and Petar Kokotovic},
  title = {Nonlinear and Adaptive Control Design},
  year = {1995},
}

@book{Spong,
  author = {Mark W. Spong and Seth Hutchinson and Mathukumalli Vidyasagar},
  title = {Robot Modeling and Control},
  edition = {2nd},
  year = {2020},
  publisher = {Wiley},
}

@article{ames,
  title = {Control barrier function based quadratic programs for safety-critical systems},
  author = {Ames, Aaron D. and Xu, Xiangru and Grizzle, Jessy W. and others},
  journal = {IEEE Trans. Autom. Control},
  volume = {62},
  number = {8},
  pages = {3861--3876},
  year = {2017},
}

@book{boyd1994linear,
  title = {Linear Matrix Inequalities in System and Control Theory},
  author = {Stephen Boyd and Lieven Vandenberghe},
  year = {1994},
  publisher = {SIAM},
  % isbn = {978-0898713238}
}

@book{freeman2008robust,
  title = {Robust nonlinear control design: State-Space and Lyapunov techniques},
  author = {Randy Freeman and Petar V. Kokotovic},
  year = {2008},
  publisher = {Birkhäuser},
  % isbn = {978-0817645652}
}

\end{document}